\documentclass[runningheads]{svmult}

\usepackage{makeidx}   
\usepackage{graphicx}  
\usepackage{subeqnar}  
\usepackage{multicol}  
\usepackage{physmubb}  
\makeindex             

\newcommand{\ffas}{\hbox{$\,.\!\!^{\prime\prime}$}}
\newcommand{\ffs}{\hbox{$\,.\!\!^{\rm s}$}}  

\usepackage{psfig}

\begin{document}
\title*{GRAVITATIONAL LENSING AT MILLIMETER WAVELENGTHS}
\toctitle{Gravitational Lensing at Millimeter Wavelengths}
%
%
\titlerunning{Gravitational Lensing at Millimeter Wavelengths}
%
\author{Tommy Wiklind\inst{1} \and Danielle Alloin\inst{2}}

\authorrunning{Wiklind \& Alloin}
%
%

\institute{
Dept. of Astronomy \& Astrophysics, Onsala Space Observatory, SE-43992 Onsala, Sweden
\and 
European Southern Observatory, Casilla 19001 Santiago 19, Chile}

\maketitle              

\begin{abstract}
The study of gas and dust at high redshift gives an unbiased view of
star formation in obscured objects as well as the chemical evolution
history of galaxies.  With today's millimeter and submillimeter
instruments observers use gravitational lensing mostly as a tool to
boost the sensitivity when observing distant objects.  This is evident
through the dominance of gravitationally lensed objects among those
detected in CO rotational lines at $z>1$. It is also evident in the
use of lensing magnification by galaxy clusters in order to reach
faint submm/mm continuum sources.  There are, however, a few cases
where millimeter lines have been directly involved in understanding
lensing configurations.  Future mm/submm instruments, such as the ALMA
interferometer, will have both the sensitivity and the angular
resolution to allow detailed observations of gravitational lenses.
The almost constant sensitivity to dust emission over the redshift
range $z \approx 1-10$ means that the likelihood for strong lensing of
dust continuum sources is much higher than for optically selected
sources. A large number of new strong lenses are therefore likely to
be discovered with ALMA, allowing a direct assessment of cosmological
parameters through lens statistics.  Combined with an angular
resolution $<0\ffas1$, ALMA will also be efficient for probing the
gravitational potential of galaxy clusters, where we will be able to
study both the sources and the lenses themselves, free of obscuration
and extinction corrections, derive rotation curves for the lenses,
their orientation and, thus, greatly constrain lens models.
\end{abstract}

\section{Introduction}

Rapid progress in the development of millimeter astronomical
facilities, such as the increase of antennae sizes and/or of 
the number of array elements, or as the continuing improvement 
in the sensitivity of detectors, have now made it possible to
explore the high redshift universe in this window and therefore
to exploit the potentialities of gravitational lensing effects.

Why is it so important to explore this wavelength domain? In 
one short statement: the presence of cold dust and of molecular
material can be traced in this window and both components witness 
the formation of heavy elements. If they are detected in galaxies 
at high redshifts, they allow us to probe star formation in the 
early universe. They reveal as well processes related to the 
startup of active galactic nuclei (AGN) and signal the presence 
of massive black-holes. Large amounts of molecular gas are 
encountered in the close environment of the central engine in 
AGN. This material is often regarded as the fuel which allows 
to activate the AGN. An evolutionary scenario would then connect 
IR-luminous galaxies rich in molecular material and with intense 
star formation to the formation/feeding of massive black holes. 

Several fundamental questions are therefore underlying the
search for dust and molecular gas at high redshift: the redshift
of galaxy formation? the chemical evolution of the universe with
time? the evolution of the dust content in the universe at
early ages? the epoch and the scenario of the formation of
massive black holes? the startup and evolution of AGN activity?
Do we have already some clues to answer these questions? The most
powerful AGNs, quasars, are now detected up to redshift around 6.5,
and galaxies up to redshift 6. So we know that in this redshift
range the universe already hosted galaxies and massive black holes
and that its metal content was substantial since the spectra of high
redshift AGNs are very similar to those of low redshift AGNs. Yet,
only a few objects are known at these high redshifts and this may 
provide a biased view. It is therefore mandatory to enlarge the 
sample and of course, the goal is also to push the redshift limit.

Pushing the redshift limit also means that we are investigating  
sources with lower and lower flux density. This can be achieved
through technical improvements, using larger collectors and better
detectors. The ALMA project is showing the way. Another manner is
to take advantage of the effects induced by gravitational lensing
(for a review of its theoretical basis, see the comprehensive
book by Schneider et al.~\cite{schneider92}). Firstly, image
magnification allows us to detect more distant sources of cold
dust and molecular gas of a given intrinsic luminosity, or to
detect at a given redshift sources of fainter intrinsic luminosity.
The latter in particular is important for good determinations of
luminosity functions. Secondly, differential magnification effects
can be used as an elegant tool to probe the size of molecular and
dusty structures in the lensed source, as long as the lensing
system provides the appropriate geometry. In this case, it is
imperative to have an excellent model of the lensing system,
as any structural information about the source itself for example
is recovered by tracing the image back through the lensing system.
Both aspects will be discussed at length in this paper. In some
cases molecular absorption lines allow us
to obtain information about the lensing galaxy itself.

Apart from hydrogen and helium, carbon and oxygen are the heavy
elements with highest abundance in the universe. Therefore, the 
CO molecule is the most suitable candidate for the detection of
molecular gas in emission at high redshift. The CO molecules can
be detected directly through their thermal line emission in the
source or as silhouetted absorbers along the line of sight to a
background source. The latter may occur for example for the
lensing galaxy. Several other molecules have been detected at
high redshift, HCO$^+$, HCN, H$_2$CO..., while dust is detected
essentially through its thermal emission.

\medskip

We provide in Sect.~\ref{molemission} an overview of the CO line
emission and of the high redshift CO sources detected so far.
The role played by gravitational lensing in studying CO sources at
high redshift is highlighted. In Sect.~\ref{molabs0} we review
molecular absorption and the importance of such measurements to
investigate the properties of the lensing galaxies.
Sect.~\ref{dustcont} introduces the dust continuum emission and
the use of differential magnification effects which can be made
to probe the dust content of the lensed objects. Three cases
particularly well studied are discussed in detail in
Sect.~\ref{casestudies}. In Sect.~\ref{pks1830model} existing lens
models for PKS1830-211 are reviewed and a new one introduced. Finally,
in Sect.~\ref{futureprosp} the future of this type of investigations
is presented in the perspective of new instrumental developments in
general and of ALMA in particular.

\section{MOLECULAR EMISSION} \label{molemission}

The goal of this section is primarily to highlight the benefits of
exploiting gravitational lensing effects in the millimeter range, 
that is in CO line emission. Therefore, after some brief comments
on the pioneering observations of low redshift sources, we shall
concentrate on the results obtained on high redshift sources.
This section deals essentially with the CO line emission, while
the following sections will discuss molecular absorption lines
and dust thermal emission.

\bigskip

Detection and measurements of the $^{12}$CO rotational transitions in 
Galactic and extragalactic sources have had a great impact on the 
development of astrochemistry. The J=1-0 CO transition is excited
by collisions with H$_2$ molecules, even in clouds at low kinetic 
temperature. The low-J transitions are in general optically thick, 
the opacity being determined observationally by the relative 
intensity of the corresponding $^{13}$CO transition. On the contrary,
the high-J transitions are optically thin. Therefore, it is quite
interesting to perform multi-transition studies to ascertain in 
a more secure fashion the physical conditions, temperature and
density, in the emitting molecular material. This is particularly
true in the case of very dense molecular material exposed to 
intense radiation fields, like in the environment of an AGN or in
powerful star forming regions. The conversion factor which is used
to derive the total mass of molecular material from the observed
L$_{\rm CO}$, is also highly dependent on the physical  conditions
in the molecular material. It has been determined to be
4.6 M$_{\odot}$ (K km s$^{-1}$ pc$^2$)$^{-1}$ in standard Milky Way
clouds \cite{solomon87}. Recently it has been shown that lower
values of the conversion factor (by up to a factor 10) should be
used in the case of dense and warm material as usually encountered
in a molecular torus around an AGN \cite{downes98}.

\subsection{Low- and intermediate redshift galaxies}
Following far-infrared (FIR) observations by IRAS, an important
population of IR-luminous (dust-rich) galaxies was found.
This was the starting point for investigating as well their
molecular content, using the millimeter facilities available in
the late 80's. A number of IR-luminous galaxies and AGNs were
detected, mostly in the CO(1-0) transition and the field
develop quickly. Regarding low redshift sources, let us briefly
mention the detection of AGNs such as Mrk 231 \cite{sanders87},
Mrk 1014 \cite{sanders88}, I Zw 1 \cite{barvainis89}
and of some low redshift quasars \cite{sanders89}
\cite{alloin92} or radio galaxies \cite{phillips87}
\cite{mirabel89} \cite{mazzarella93}. At moderate
redshift, CO was detected in the radio galaxy 3C48 ($z=0.369$)
\cite{scoville93}. This search is continuing through e.g. the
Caltech CO high and low redshift radio galaxy survey \cite{evans99}.

\medskip

On the side of CO sources at high redshift (z$>$2), the first object
detected was IRAS 10214+4724 \cite{brown92} \cite{solomon92}.
All along the 90's, millimeter dishes and interferometer
arrays in service were pushed to their limits in searching for
other candidates, selected for example upon the strength of their
submillimeter flux. A large amount of observing time was dedicated
to such programs at the OVRO, BIMA, Nobeyama and IRAM facilities. At 
face value, the success rate in detecting high redshift CO sources 
has been modest. One reason for this is the uncertainty in the 
precise redshift of the emitting molecular gas under search, while the 
backends of the instruments are narrow in comparison to the redshift 
range to be explored. Another reason is of course the limited
sensitivity of existing instruments. Only the most luminous and
most gas-rich systems can be detected. The situation should improve
with new facilities such as ALMA (See Sect.~\ref{future}).
Sources detected so far are detailed below, in order of increasing
redshift.

\begin{table}
\caption{Galaxies at z$>$1 with molecular emission (January 2002)}
\begin{center}
\begin{tabular}{lccccccc}
\ \\
\hline
\ \\
Name & z &\hspace{0.75cm} & $M_{\rm H_2}$ & $M_{\rm d}$/M$_{\odot}$ & $L_{\rm FIR}$/L$_{\odot}$ & Grav. lens & Ref. \\
\ \\
\hline
\ \\
BR1202$-$0725   &\ 4.69 && $6 \times 10^{10}$ & $2 \times 10^{8}$  & $\sim 1 \times 10^{12}$ & ?  &
\cite{ohta96}, \cite{omont96} \\
\ \\
BR0952$-$0115   &\ 4.43 && $3 \times 10^{9}$  & $3 \times 10^{7}$  & $\sim 1 \times 10^{12}$ & YES &
\cite{guilloteau99} \\
\ \\
BRI1335$-$0414  &\ 4.41 && $1 \times 10^{11}$ & continuum det.     & ---                     & NO? &
\cite{guilloteau97} \\
\ \\
PSS2322+1944    &\ 4.12 && $3 \times 10^{11}$ & $1 \times 10^{9}$  & ---                     & ?   &
\cite{cox02} \\
\ \\
\hline
\ \\
APM08279+$$5255 &\ 3.91 && $2 \times 10^{9}$  & $1 \times 10^{7}$  & $\sim 8 \times 10^{13}$ & YES &
\cite{downes99b} \\
\ \\
4C60.07         &\ 3.79 && $8 \times 10^{10}$ & $2 \times 10^{8}$  & $\sim 2 \times 10^{13}$ & NO  &
\cite{papadopoulos00} \\
\ \\
6C1909$+$722    &\ 3.53 && $4 \times 10^{10}$ & $2 \times 10^{8}$  & $\sim 2 \times 10^{13}$ & NO  &
\cite{papadopoulos00} \\
\ \\
MG0751+2716     &\ 3.20 && $8 \times 10^{10}$ & ---                & ---                     & YES &
\cite{barvainis02} \\
\ \\
\hline
\ \\
SMM02399$-$0136 &\ 2.81 && $8 \times 10^{10}$ & continuum det.     & $\sim 1 \times 10^{13}$ & YES &
\cite{frayer98} \\
\ \\
MG0414$+$0534   &\ 2.64 && $5 \times 10^{10}$ & continuum det.     & ---                     & YES &
\cite{barvainis98} \\
\ \\
SMM14011$+$0252 &\ 2.56 && $5 \times 10^{10}$ & continuum det.     & $\sim 3 \times 10^{12}$ & YES &
\cite{frayer99} \\
\ \\
H1413$+$117     &\ 2.56 && $2 \times 10^{9}$  & $1 \times 10^{8}$  & $\sim 2 \times 10^{12}$ & YES &
\cite{barvainis94} \\
\ \\
53W002          &\ 2.39 && $1 \times 10^{10}$  & weak continuum    & ---                     & NO  &
\cite{scoville97} \\
\ \\
F10214$+$4724   &\ 2.28 && $2 \times 10^{10}$  & $9 \times 10^{8}$ & $\sim 7 \times 10^{12}$ & YES &
\cite{brown92} \\
\ \\
\hline
\ \\
HR 10           &\ 1.44 && $7 \times 10^{10}$  & $2 \times 10^{8}$ & $\sim 9 \times 10^{11}$ & NO  &
\cite{andreani00} \\
\ \\
\hline
\end{tabular}
\ \\
\ \\
\end{center}
Masses and fluxes corrected for gravitational magnification (approx.) \\
\ \\
\label{cotable}
\end{table}

\subsection{High redshift galaxies}
The first source discovered, IRAS 10214+4724, at $z=2.285$, has been 
detected in CO(3-2), CO(4-3) and CO(6-5) \cite{brown92} 
\cite{solomon92}. A report on the detection of CO(1-0) \cite{tsuboi94}
remains to be confirmed. The source is a gravitationally 
lensed ultraluminous IR-galaxy \cite{broadhurst95} \cite{downes95}.
The magnification factor is found to be around 10 for the 
CO source which has an intrinsic radius of 400 pc. Conversely, the 
far-IR emission detected in this object is magnified 13 times and 
arises from a source with radius 250 pc, while the mid-IR is 
magnified 50 times and arises from a source with radius 40 pc. 
After correcting for magnification, and using a conversion factor 
L$^{\prime}_{\rm CO}$ to M(H$_2$) of 4 M$_{\odot}$
(K km s$^{-1}$ pc$^2$)$^{-1}$ \cite{radford91}, the molecular
gas mass is found to be $2 \times 10^{10}$ M$_{\odot}$, in agreement
with the estimated dynamical mass $3 \times 10^{10}$ M$_{\odot}$.
As noted above however, such a value for the conversion factor,
obtained from CO(1-0) observations of Galactic molecular clouds,
might not be applicable in the case of warmer and denser molecular
material. A value for the conversion factor which is 5 times lower
than the standard has been found in a study of extreme starbursts
in IR-luminous galaxies \cite{downes98}.
Hence, the mass value quoted above should regarded as an upper limit
to the mass of the molecular gas in IRAS 10214+472. Still pending is
the question of the CO line emission share between a hidden AGN and
a starburst in the 400 pc region surrounding the AGN. We notice also
that the large extension (3 to 12 kpc) in CO emission reported by
\cite{scoville95} has not been confirmed by the IRAM interferometer
data \cite{downes95}.

\medskip

One interesting case is that of the radio galaxy 53W002 at $z=2.394$,
located at the center of a group of $\sim$20 Lyman-$\alpha$ emitters.
A possible detection of the CO(1-0) line at Nobeyama was reported in
\cite{yamada95}, although not yet confirmed by others. The first
detection in CO(3-2) by OVRO \cite{scoville97} suggested a
large extension (30 kpc) and the existence of a velocity gradient.
None of these features has been confirmed by an IRAM interferometer
data set with higher signal to noise ratio \cite{alloin00}.
From an astrometric analysis it is found that the 8.4 GHz and CO
source are coincident, at a location consistent with that of the
optical/UV continuum source. The most likely origin of the molecular
emission is therefore from the close environment of the AGN.
One should notice that 53W002 is definitely not a gravitationally 
lensed source. Using a conversion factor L$^{\prime}_{\rm CO}$ to
M(H$_2$) in the range 0.4 - 0.8 M$_{\odot}$ (K km s$^{-1}$ pc$^2$)$^{-1}$,
more appropriate for dense and warm molecular gas around an AGN
\cite{barvainis97}, the resultant molecular gas mass is found
to be in the range $(0.6 - 1.0) \times 10^{10}$ M$_{\odot}$.

\medskip

The Cloverleaf, H1413+117 is a well known gravitationally lensed
Broad Absorption Line (BAL) quasar at $z=2.558$ \cite{magain88}.
Its CO(3-2) transition
was first observed with the IRAM 30m dish \cite{barvainis94}
and then with BIMA \cite{wilner95}. Later, the CO(4-3), 
CO(5-4) and CO(7-6) transitions have been detected, together with 
HCN(4-3) and a fine-structure line of CI. From a detailed analysis 
of these transitions, the molecular gas was found to be warm and 
dense \cite{barvainis97} with a low conversion factor of
0.4 M$_{\odot}$ (K km s$^{-1}$ pc$^2$)$^{-1}$. High resolution maps
in the CO(7-6) transition were obtained with OVRO \cite{yun97}
and with the IRAM interferometer \cite{alloin97} \cite{kneib98a}.
The IRAM map has the best resolution and signal to noise ratio.
Comparing with the HST images and exploiting differential gravitational
effects, the CO(7-6) map allowed a derivation of both the the size and
the kinematics of the molecular/dusty torus around the quasar central
engine \cite{kneib98a} (for further details see Sect.~\ref{cloverleaf}).
After correcting for the amplification factor (30 according to the model
used for the lensing system), and using the mean conversion factor
0.6 M$_{\odot}$ (K km s$^{-1}$ pc$^2$)$^{-1}$ (derived by Barvainis et
al.~\cite{barvainis97}), the derived mass of molecular gas M(H$_2$) is
$2 \times 10^9$ M$_{\odot}$, in agreement with the dynamical mass of
$8 \times 10^8$ M$_{\odot}$.

\medskip

The source SMM 14011+0252 was observed with OVRO in CO(3-2) at 
$z=2.565$ \cite{frayer99}, following its discovery as a strong
submillimeter source detected in the course a survey of rich lensing
clusters \cite{smail98b}. Correcting for an amplification factor
of 2.75 and using a conversion factor of 4
M$_{\odot}$ (K km s$^{-1}$ pc$^2$)$^{-1}$, the mass of molecular gas
turns out to be $5 \times 10^{10}$ M$_{\odot}$, while the dynamical
mass is found to be larger than $1.5 \times 10^{10}$ M$_{\odot}$.
The CO emission is extended on scales of $\sim$10 kpc and associated
with likewise extended radio continuum emission \cite{ivison01}.
Optical and near-infrared (NIR) imaging shows two objects (designated
J1 and J2), separated by 2\ffas1 \cite{ivison00}. Both the
molecular gas and the radio continuum, however, have their strongest
emission $\sim$1$^{\prime\prime}$ north of the J1/J2 components and
are extended between J1 and J2. This suggests that the optical/NIR
emission comes from two `windows' in the obscuring molecular gas
and dust and that J1/J2 represent emission from a coherent large
galaxy. The extended nature of the radio continuum, the lack of
X-ray emission \cite{fabian00} and the lack of optical broad
emission lines (Wiklind et al. 2002 in prep) suggest that only star
formation powers the large FIR luminosity. Assuming a Salpeter initial
mass function and correcting for the gravitational magnification,
the FIR luminosity indicates a star formation rate exceeding
10$^3$ M$_{\odot}$ yr$^{-1}$.

\medskip

The gravitationally lensed quasar MG 0414+0534 was observed with 
the IRAM interferometer in the CO(3-2) transition at $z=2.639$
\cite{barvainis98}. The lensed nature of this system is known 
from a 5 GHz map \cite{hewitt92}. It displays four quasar-spots 
separated at most by $\sim 2^{\prime\prime}$. The beam of the IRAM data
(2\ffas0 $\times$ 0\ffas9) does not allow to separate the components
in the CO(3-2) velocity-integrated map. However, by fitting the UV
data directly, it has been possible to resolve the combined A components
(A1+A2) from component B and to get separate CO(3-2) spectra. The relative
strength A:B  in the 5 GHz radio continuum is 5:1. The millimeter
continuum rather shows a ratio 7:1 and differences are seen
between the A and B CO(3-2) spectra, suggesting that differential
magnification effects may be at work. The magnification factor is
unknown: hence only an upper limit can be derived for the molecular
gas mass. Assuming in addition a conservative conversion factor 
of 4 M$_{\odot}$ (K km s$^{-1}$ pc$^2$)$^{-1}$, the upper limit found for
M(H$_2$) is $2.2 \times 10^{11}$ M$_{\odot}$. This figure is below the
upper limit derived for the dynamical mass, $9 \times 10^{11}$ M$_{\odot}$.

\medskip

Another source detected first through submillimeter observations is
SMM 02399-0136 \cite{ivison98}. It is known to be gravitationally
amplified by a foreground cluster of galaxies, the amplification
factor being 2.5. This source has been detected with OVRO in 
the CO(3-2) transition at $z=2.808$ \cite{frayer98}. The mass of 
molecular gas deduced in this object, correcting for a 2.5 
amplification factor and using the conversion factor applicable to 
Galactic clouds, 4 M$_{\odot}$ (K km s$^{-1}$ pc$^2$)$^{-1}$, is
$8 \times 10^{10}$ M$_{\odot}$. From the upper limit on the apparent
size of the CO emitting source (5$^{\prime\prime}$) and the width of
the CO line (710 km s$^{-1}$), an upper limit to the mass of molecular
gas of $1.5 \times 10^{10}$ M$_{\odot}$ can be derived \cite{frayer98}.
The SCUBA results indicate that SMM 02399-0136 is an IR-hyperluminous
galaxy. On the other hand, optical data shows clearly than it hosts
as well a dust-enshrouded AGN \cite{ivison98}. A precise share of CO
emission between the two components remains to be investigated.

\medskip

In the course of a systematic CO emission survey of gravitationally 
lensed sources with the IRAM interferometer \cite{barvainis02}, 
the source MG 0751+2716 has been detected in the CO(4-3) transition
at $z=3.200$. This source was first discovered to be a gravitationally
lensed quasar, from VLA maps \cite{lehar97}. It shows four
quasar-spots with maximum separation of 0\ffas9. The lensing galaxy, 
which provides the image geometrical configuration, is part of a
group of galaxies adding another shear to the lens-system \cite{tonry99}.
The lensing system remains to be modeled in detail.
Therefore, the amplification factor is not known. However, given the
observed strength of the CO line emission it should be large. 
Assuming that the CO emission in this source is mostly from the 
close environment of the AGN, we consider a conversion factor in the 
range 0.4 to 1 M$_{\odot}$ (K km s$^{-1}$ pc$^2$)$^{-1}$.
The corresponding upper limit (no correction applied for the unknown
amplification factor) for the mass of molecular gas is in the range
$8 \times 10^{10}$ to $2 \times 10^{11}$ M$_{\odot}$ \cite{barvainis02}.

\medskip

The distant powerful radio galaxy 6C\,1909+722 has been detected in 
the CO(4-3) line at $z=3.53$, with the IRAM interferometer, and in 
dust submillimeter emission using SCUBA \cite{papadopoulos00}. 
It is unlikely to be a gravitationally lensed object. Hence, the 
derived mass of molecular material is quite large, even assuming a
conservative value for the conversion factor (about one fifth the
value derived from Galactic molecular clouds). It is found to be
in the range $(0.5-1.0) \times 10^{11}$ M$_{\odot}$.

\medskip

Another possibly unlensed powerful radio galaxy 4C\,60.07, has been 
detected in the CO(4-3) line emission at $z=3.79$ at IRAM, and in 
dust thermal emission at submillimeter wavelengths with SCUBA and 
at millimeter wavelengths at IRAM \cite{papadopoulos00}. 
Remarkably, the CO line emission extends over 30 kpc and breaks 
into two components: one corresponding to the AGN (radio core) and
a second one which is rather related to a major episode of star
formation. This state of merging is speculatively interpreted as the
formative stage of an elliptical host around the residing AGN. Again, 
the molecular mass is found to be quite large, around
$10^{11}$ M$_{\odot}$.    

\medskip
  
The gravitationally lensed BAL quasar, APM 08279+5255, has been
detected in the CO(4-3) and CO(9-8) transitions at $z=3.911$, with the
IRAM interferometer \cite{downes99b}. The CO line ratio points
towards warm and dense molecular gas. Thermal emission from the dust
component is also measured. Both the molecular and dust luminosities
appear to be very high. Gravitational amplification is therefore
suspected and has subsequently been confirmed through the detection
of three optical/NIR components \cite{ledoux98} \cite{egami00}
(see also Sect.~\ref{apm08279} and Fig.~\ref{apmfig3}). The magnification
factors for the molecular gas and dust where estimated \cite{downes99b}.
After correcting for these factors, the dust mass is found to be in the range
$(1-7) \times 10^7$ M$_{\odot}$, and the molecular gas mass in the
range $(1-6) \times 10^9$ M$_{\odot}$. In this interpretation,
the molecular/dusty component is in the form of a nuclear disk
with radius 90-270 pc orbiting the central engine of the BAL quasar.
This source looks therefore quite similar to the Cloverleaf. Recently,
however, extended low-excitation CO emission (the J=1-0 and J=2-1
transitions) have been detected using the VLA \cite{papadopoulos01}.
This extended emission is likely to be associated with a cooler molecular
component than the CO(9-8) emission.

\bigskip

\noindent
Finally, let's discuss the four sources detected so far at z larger
than 4:

\medskip

\noindent
{\bf PSS\,2322+1944}\ 
The radio quiet quasar PSS\,2322+1944 has recently been detected in 
the CO(5-4) and CO(4-3) transitions at a redshift of $z=4.12$ with
the IRAM interferometer \cite{cox02}. The velocity-inte\-gra\-ted 
CO line fluxes are $3.74\pm0.56$ and $\rm 4.24\pm0.33$ Jy km s$^{-1}$, 
with a linewidth $\approx 330$\,km s$^{-1}$. The 1.35~mm 
(250$\rm \mu m$ restwavelength) dust continuum flux density is 
7.5~mJy, in agreement with previous measurements at 1.25~mm at the 
IRAM 30m telescope \cite{omont01}, and corresponds to a dust mass
of $\approx  10^9$ M$_{\odot}$. With the present angular resolution of
the observations, no evidence for extended emission has been found yet. The
implied gas mass is estimated to be $\rm \approx 3 \times 10^{11}$
M$_\odot$, using a conversion factor of $\rm 4.6$\,M$_{\odot}$ K km s$^{-1}$
pc$^2$. The properties of PSS~2322+1944 are described in detail in \cite{cox02}.

\medskip

\noindent
{\bf BRI\,1335-0415}\ 
BRI\,1335-0415 was detected in the CO(5-4) transition with the IRAM
interferometer at $z=4.407$ \cite{guilloteau97}. The source does 
not exhibit a noticeable extension neither in the 1.35mm continuum 
nor in the CO line emission. In addition, there is no obvious sign of
gravitational lensing on the line-of-sight to this source. The authors 
have derived a very large mass of molecular gas, close to 10$^{11}$ M$_{\odot}$. 
Even with a conversion factor 3 times smaller, more appropriate for
this type of object, the mass remains a few 10$^{10}$ M$_{\odot}$.

\medskip

\noindent
{\bf BR\,0952-0115}\ 
The gravitationally lensed radio quiet quasar BR\,0952-0115 has been
detected in CO(5-4) with IRAM facilities at $z=4.43$ \cite{guilloteau99}.
A tentative estimate of the mass of molecular material
M(H$_2$) is $3 \times 10^9$ M$_{\odot}$. Note however that a more
precise model of the lens has still to be worked out.

\medskip

\noindent
{\bf BR\,1202-01215}\ 
BR\,1202-01215 is the most distant source detected in CO. It has been 
reported in the CO(5-4) transition observed with the Nobeyama array 
\cite{ohta96}, and in the CO(4-3), CO(5-4) and CO(7-6) lines
observed with IRAM facilities \cite{omont96}. The CO maps show
two separate sources on the sky: one is coincident with the optical
quasar (for this source CO(5-4) provides $z=4.695$), while the other is 
located 4'' to the North-West, where no optical counterpart is found 
(for this source CO(5-4) provides $z=4.692$). At this redshift the
4$^{\prime\prime}$ extension corresponds to a de-projected distance
of 12-30 kpc. It is uncertain whether this object is gravitationally
lensed or not. Is the North-West source a second image of the quasar?
There are hints that this might be the case as a strong gravitational
shear has been measured in the field (Fort and D'Odorico, private
communication). Else, each of two separate sources ought to have its
own heating source, AGN or starburst. Assuming no gravitational boost
and using a conversion factor of 4 M$_{\odot}$ (K km s$^{-1}$ pc$^2$)$^{-1}$,
the mass of molecular gas is quoted to be $6 \times 10^{10}$ M$_{\odot}$.

\bigskip

A summary of the sources properties is provided in Table~\ref{cotable}.
This quick compilation of high redshift CO sources has prompted a 
number of key-issues:

\bigskip

\noindent
{\it (i)}\ 
A major difficulty in detecting high redshift CO sources is the 
lack of precision in our guess for the redshift of the molecular gas. 
The instantaneous frequency coverage of current backends requires that
the redshift is known a-priori with a precision of a few percent.
Why is this condition hard to fulfill? The molecular gas emission in
distant objects can arise from the close environment of an AGN. Yet,
published redshifts for distant AGN are mostly measured from emission
lines of highly ionized species which can be strongly affected by 
winds. Indeed velocity offsets of up to 2500 km s$^{-1}$ have been
observed between the CO lines and the blue-shifted CIV line for example
(e.g.~\cite{downes99b}). Hopefully, this limitation will be overcome
with the next generation of backends.

\medskip

\noindent
{\it (ii)}\ An uncertain part in the interpretation of the observed CO line 
intensities lies with the physical state of the molecular gas and the
conversion factor L$^{\prime}_{\rm CO}$ to M(H$_2$) to be applied in
the case of high redshift sources. When several CO transitions are observed
(such as for IRAS10214+4724, H1413+117, APM\,08279+5255 and BR\,1202-0725),
the physical conditions of the molecular gas can be analyzed, pointing
towards warm (T$\sim$100 K), dense (a few 10$^3$ cm$^{-3}$) and moderately
optically thick material. Such conditions could very well characterize
molecular gas in the proximity of an AGN. Conversely, the conditions
of the molecular gas in an extended starburst may be more similar to
those encountered in Galactic molecular clouds. The conversion factor
depends on the physical conditions of the molecular gas. It ranges from
a value of 4 M$_{\odot}$ (K km s$^{-1}$ pc$^2$)$^{-1}$ in Galactic clouds
\cite{radford91}, to 1 M$_{\odot}$ (K km s$^{-1}$ pc$^2$)$^{-1}$ in 
IR-ultraluminous galaxies \cite{downes98} and possibly 0.4
M$_{\odot}$ (K km s$^{-1}$ pc$^2$)$^{-1}$ in the surroundings of an AGN
\cite{barvainis97}. Therefore, it would be important to have some
clues about the share AGN/starburst in the heating mechanism for high
redshift CO sources. In that respect, the CO line width and the
compactness of the source may bring some pieces of information.

\medskip

\noindent
{\it (iii)}\ Finally, the outmost efforts should be made to find out
whether a  source is gravitationally lensed or not, before the claim
for the presence of a huge amount of molecular gas (10$^{11}$ M$_{\odot}$)
can be taken as a starting point for modeling. Weak shear from an
intervening galaxy cluster, like in the cases of SMM\,14011+0252 and
SMM\,02399-0136, induces a mild magnification factor in the range 2-3.
Strong shear (possibly combined with weak shear), induces magnification
factors of up to 30! This would decrease by one order of magnitude the
mass of molecular gas derived. If, at the same time, the applicable
conversion factor is on the low side of its possible values range, a
reduction of the actual molecular gas mass by another order of magnitude
would apply. 

\bigskip

In conclusion, it is important to search for other high redshift CO
sources and, at the same time, to investigate carefully the nature of
their environment/line-of-sight and the physical conditions in their 
molecular gas.

\section{MOLECULAR ABSORPTION LINES} \label{molabs0}

Another method to study molecular gas at high redshift is to
observe molecular rotational lines in absorption rather than emission.
Whereas emission is biased in favor of warm and dense molecular
gas, tracing regions of active high mass star formation, molecular
absorption lines trace excitationally cold gas. This is important since
a large part of the molecular gas mass may reside in regions far away
from massive star formation and therefore remain largely unobserved in
emission.

Molecular absorption occurs whenever the line of sight to a background
quasar passes through a sufficiently dense molecular cloud. In
contrast to optical absorption lines seen towards most high redshift
QSOs, molecular absorption is invariably associated with galaxies,
either in the host galaxy of the continuum source or along the line of
sight. In nearby galaxies molecular gas is strongly
concentrated to the central regions, making the likelihood for
absorption largest whenever the line of sight passes close to
the center of an intervening galaxy. This, of course, means that
molecular absorption in intervening galaxies is likely to be
associated with gravitational lensing, and vice versa. Indeed,
the only known systems of intervening absorption (B0218+357 and
PKS1830-211) are gravitationally lensed and the absorption probes
molecular gas in the lensing galaxy.
Molecular absorption lines can thus be used to study the neutral
and dense interstellar medium in lenses. At the present the sample
of lens galaxies probed by molecular absorption lines is limited, but
with the advent of a new sensitive interferometer instrument like
ALMA, the number of potential candidate systems will increase
substantially and make it possible to probe the molecular interstellar
medium in the lens galaxies in some detail.
Moreover, the molecular absorption lines provide unique kinematical
information which is valuable when constructing a model of the lensed
system.

\subsection{Detectability} \label{molabs1}

As mentioned above, molecular absorption traces a different gas
component then emission lines. For optically thin {\it emission}
the integrated signal $I_{\rm CO}$ is
$$
I_{\rm CO} = \int{T_{\rm a}\,dv}\  \propto\ 
N_{\rm tot}\,T^{-1}\,e^{-E_{\rm u}/kT}\,(e^{{\rm h}\nu/{\rm k}T} - 1)
\left[J(T) - J(T_{\rm bg})\right]\ \ ,
$$
where $N_{\rm tot}$ is the total column density of a given molecular
species, $E_{\rm u}$ the upper energy level of a transition with
$\Delta E = h\nu$, $T_{\rm bg}$ is the local temperature of the Cosmic
Microwave Background Radiation (CMBR) and
$J(T) = (h\nu/k)(e^{h\nu/kT} - 1)^{-1}$.
When $T \rightarrow T_{\rm bg}$ all the molecules reside in the ground
rotational state $J=0$ and the signal disappears.
For molecular {\it absorption} the observable is the velocity integrated
opacity $I_{\tau_{\nu}}$:
$$
I_{\tau_{\nu}} = \int{\tau_{\nu}\,dv}\  \propto\ 
N_{\rm tot}\,T^{-1}\,\mu_0^2\,e^{-E_l/kT}\,(1 - e^{-h\nu/kT}) \ \ ,
$$
where $N_{\rm tot}$ is again the total column density of a given molecular
species, while $E_l$ is now the lower energy level and $\mu_0$ is the
permanent dipole moment of the molecule. For the ground
transition\footnote{This expression is strictly speaking only true for
linear molecules.},
$E_l = 0$,
$I_{\tau_{\nu}} \propto N_{\rm tot}\,T^{-1}\,\mu_0^2\,(1 - e^{-h\nu/kT})
\approx (h\nu/k)\,N_{\rm tot}\,\mu_0^2\,T^{-2}$.
In contrast to emission lines, the observed integrated opacity increases
as the temperature $T$ decreases.

Molecules are generally excited through collisions with molecular
hydrogen H$_2$. The excitation temperature, $T_{\rm x}$, therefore depends
strongly on the H$_2$ density. The collisional excitation is balanced
by radiative decay and a steady-state situation with $T_{\rm x} = T_{\rm k}$
requires a certain critical H$_2$ density. For CO, which has a small
permanent dipole moment $\mu_0$, the critical density is rather low,
$4 \times 10^4$ cm$^{-3}$, while molecules with higher dipole moments
require higher densities. For instance, HCO$^+$ which has a dipole moment
more than 30 times larger than that of CO, the critical density is
$2 \times 10^7$ cm$^{-3}$.

\medskip

\begin{figure}[h]
\psfig{figure=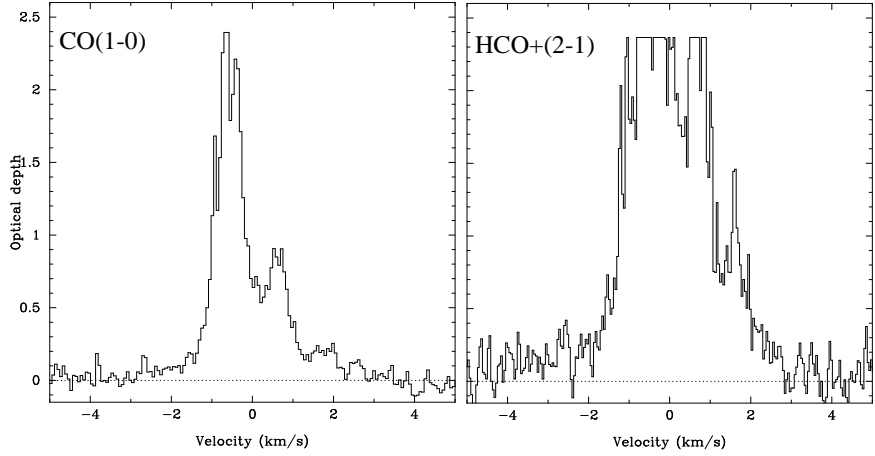,width=11.5cm,angle=-90.0}
\caption[]{Plots of the observed opacity for the CO(1-0) and
HCO$^+$(2-1) transitions seen at $z=0.25$ towards PKS1413+135.
The cut-off at opacities $> 2$ are due to saturation of the
signals. The opacity of the HCO$^+$(2-1) line is larger than
that of the CO(1-0) line despite of an abundance which is
$10^{-3} - 10^{-4}$ that of CO.}
\label{opacity}
\end{figure}

The strong dependence of the opacity on the permanent dipole moment
means that absorption preferentially probes low excitation gas, i.e.
a cold and/or diffuse molecular gas component. If multiple gas
components are present in the line of sight, with equal column densities
but characterized by different excitation temperatures, absorption
will be most sensitive to the gas component with the lowest temperature.
The dependence of the opacity on the permanent dipole moments also
means that molecules much less abundant than CO can be as easily
detectable. For instance, HCO$^+$ has an abundance which is of the
order $5 \times 10^{-4}$ that of CO, yet it is as easy, or easier,
to detect in absorption as CO. This is illustrated in
Fig.~\ref{opacity}, where the observed opacity of the CO(1-0)
and HCO$^+$(2-1) transitions at $z=0.25$ are compared. In this
particular case, the HCO$^+$ line has a higher opacity than the
CO line.

\subsection{Observables}

Analysis of the molecular absorption lines gives important
information about both the physical and chemical properties
of the interstellar medium. This can have implications for
identifying the type of galaxy causing the absorption and,
in some cases, help to identify the morphological type of
lenses. In this section a short description of the analysis
that can be done is presented. A more detailed description
can be found in the references given in the text.

\subsubsection{Optical depth.}\ 
The observed continuum temperature, $T_{\rm c}$, away from
an absorption line can be expressed as
$T_{\rm c}=f_{\rm s}J(T_{\rm b})$,
where $f_{\rm s}$ is the beam filling factor of the region
emitting continuum radiation, $T_{\rm b}$ is the brightness
temperature of the background source and
$J(T)=(h\nu/k)/[1-\exp(-h\nu/kT)]$ (e.g.~\cite{wiklind97}).
The spatial extent of the region emitting continuum radiation
at millimeter wavelengths is unknown but is certain to be smaller
than at longer wavelengths.
The BL Lac 3C446 has been observed with mm-VLBI and has a size
$<30\,\mu$arcseconds \cite{lerner93}.
Since the angular size of a single dish telescope
beam at millimeter wavelengths is typically $10^{\prime\prime}-25^{\prime\prime}$,
the brightness temperature of the background source, $T_{\rm b}$, is at
least $10^{9} \times T_{\rm c}$.
This means that the local excitation temperature of the molecular
gas is of no significance when deriving the opacity. The excitation
does enter, however, when deriving column densities.

\subsubsection{Excitation temperature and column density.}\ 
The excitation temperature, $T_{\rm x}$, relates the relative population
of two energy levels of a molecule as:
$\frac{n_{2}}{n_{1}} = \frac{g_{2}}{g_{1}}\ e^{-h\nu_{21}/kT_{\rm x}}$,
where $g_{i}$ is the statistical weight for level $i$ and
$h\nu_{21}$ is the energy difference between two rotational levels.
In order to derive $T_{\rm x}$ we must link the fractional population
in level $i$ to the total abundance. This is done by invoking the
weak LTE-approximation\footnote{In the weak LTE-approximation
$T_{\rm x} \approx T_{\rm rot}$, but the rotational temperature $T_{\rm rot}$ is
not necessarily equal to the kinetic temperature and can also be different for
different molecular species.}. We can then use the partition function
$Q(T_{\rm x})=\sum_{J=0}^{\infty}g_{J}\,e^{-E_{J}/kT_{\rm x}}$
to express the total column density, $N_{\rm tot}$, as
\begin{eqnarray} \label{opacity4}
N_{\rm tot} & = & {8\pi \over c^{3}} {\nu^{3} \over g_{J} A_{J,J+1}} f(T_{\rm x})
             \ \int \tau_{\nu} dV\ ,\nonumber \\
\ \\
f(T_{\rm x}) & = & {Q(T_{\rm x}) e^{E_{J}/kT_{\rm x}} \over
1-e^{-h\nu/kT_{\rm x}}}\ ,\nonumber
\end{eqnarray}
where $\int \tau_{\nu} dV$ is the observed optical depth integrated
over the line for a given transition, $g_{\rm J}=2J+3$ for a
transition $J \rightarrow J+1$, and $E_{\rm J}$ is the energy of
the rotational level $J$.
By taking the ratio of two observed transitions from the same molecule, the
excitation temperature can be derived.
The strong frequency dependence of the column density in Eq.~\ref{opacity4}
is only apparent since the Einstein coefficient, $A_{J,J+1}$, is
proportional to $\nu^3$.

\begin{table}[t]
\caption{Properties of molecular absorption line systems.}
\begin{center}
\renewcommand{\arraystretch}{1.4}
\setlength\tabcolsep{3pt}
\scriptsize
\begin{tabular}{lcccccrc}
\noalign{\smallskip}
\hline
\noalign{\smallskip}
 & & & & & & & \\
Source & z$_{\rm a}^{(a)}$ & z$_{\rm e}^{(b)}$ &
$N_{\rm CO}$ & $N_{\rm H_2}$ & $N_{\rm HI}^{(c)}$ &
A$_{\rm V}^{\prime\ (d)}$ & $N_{\rm HI}/N_{H_2}$ \\
 & & & cm$^{-2}$ & cm$^{-2}$ & cm$^{-2}$ & & \\
\hline\noalign{\smallskip}
 & & & & & & & \\
Cen A         & 0.00184 & 0.0018 & $1.0 \times 10^{16}$ & $2.0 \times 10^{20}$ 
& $1 \times 10^{20}$ & 50 & 0.5 \\
PKS1413+357   & 0.24671 & 0.247  & $2.3 \times 10^{16}$ & $4.6 \times 10^{20}$
& $1.3 \times 10^{21}$ & 2.0 & 2.8 \\
B3\,1504+377A & 0.67335 & 0.673  & $6.0 \times 10^{16}$ & $1.2 \times 10^{21}$
& $2.4 \times 10^{21}$ & 5.0 & 2.0 \\
B3\,1504+377B & 0.67150 & 0.673   & $2.6 \times 10^{16}$ & $5.2 \times 10^{20}$
& $<7 \times 10^{20}$ & $<$2 & $<$1.4 \\
B\,0218+357   & 0.68466 & 0.94   & $2.0 \times 10^{19}$ & $4.0 \times 10^{23}$
& $4.0 \times 10^{20}$ & 850 & $1 \times 10^{-3}$ \\
PKS1830--211A & 0.88582 & 2.507  & $2.0 \times 10^{18}$ & $4.0 \times 10^{22}$
& $5.0 \times 10^{20}$ & 100 & $1 \times 10^{-2}$ \\
PKS1830--211B & 0.88489 & 2.507  & $1.0 \times 10^{16\ (e)}$ &
$2.0 \times 10^{20}$ & $1.0 \times 10^{21}$ & 1.8 & 5.0 \\
PKS1830--211C & 0.19267 & 2.507   & $<6 \times 10^{15}$                   &
$<1 \times 10^{20}$ & $2.5 \times 10^{20}$ & $<$0.2 & $>$2.5 \\
 & & & & & & & \\
\noalign{\smallskip}
\hline
\noalign{\smallskip}
\end{tabular}
\end{center}
\ \\
$(a)$\ Redshift of absorption line. \\
$(b)$\ Redshift of background source. \\
$(c)$\ 21cm HI data taken from~\cite{carilli92} \cite{carilli93}
\cite{carilli97a} \cite{carilli97b}. A spin-temperature of 100 K
and a area covering factor of 1 was assumed. \\
$(d)$\ Extinction corrected for redshift using a Galactic extinction law. \\
$(e)$\ Estimated from the HCO$^{+}$ column density
of $1.3 \times 10^{13}$\,cm$^{-2}$. \\
\label{molabssystem}
\end{table}

\subsection{Known Molecular Absorption Line Systems} \label{molabs}

There are four known molecular absorption line systems at high redshift:
z$=$0.25-0.89. These are listed in Table~\ref{molabssystem} together
with data for the low redshift absorption system seen toward the radio
core of Centaurus A. For the high redshift systems, a total of 18 different
molecules have been detected, in 32 different transitions.
This includes several isotopic species: C$^{13}$O, C$^{18}$O, H$^{13}$CO,
H$^{13}$CN and HC$^{18}$O$^+$. As can be seen from Table~\ref{molabssystem},
the inferred H$_2$ column densities varies by $\sim 10^3$. The isotopic
species are only detectable towards the systems with the highest column
densities: B0218+357 and PKS1830-211, which are also the systems where
the absorption originates in lensing galaxies.
The large dispersion in column densities is reflected in the large spread
in optical extinction, $A_{\rm V}$, as well as the atomic to molecular ratio.
Systems with high extinction have 10-100 times higher molecular gas
fraction than those of low extinction.

\subsubsection{Absorption in the host galaxy.}\ 
Two of the four known molecular absorption line systems are situated within
the host galaxy to the `background' continuum source: PKS1413+135 \cite{wiklind94}
and B3\,1504+377 \cite{wiklind96b}. The latter exhibits two absorption line
systems with similar redshifts, z$=$0.67150 and 0.67335. The separation
in restframe velocity is 330 km\,s$^{-1}$. This is the type of signature
one would expect from absorption occurring in a galaxy acting as a gravitational
lens, where the line of sight to the images penetrate the lensing galaxy on
opposite sides of the galactic center.
However, in this case, as well as for PKS1413+135, high angular
resolution VLBI images show no image multiplicity, despite impact parameters
less than 0\ffas1 (e.g.~\cite{perlman96} \cite{xu95}).
The continuum source must therefore be situated within or very near the
obscuring galaxy.

\begin{figure}
\psfig{figure=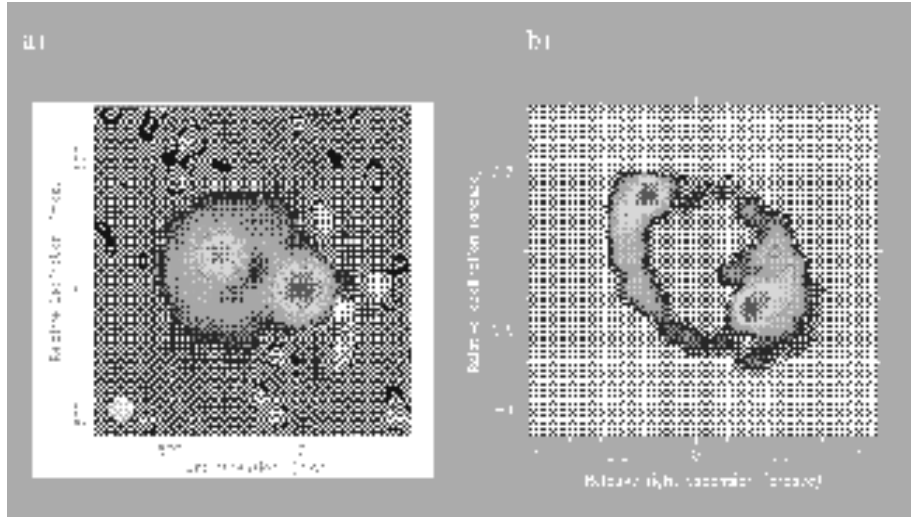,width=12.0cm,angle=0.0}
\caption[]{{\bf a)}\ A 15 GHz radio image of the gravitational lens
B0218+357 obtained with the VLA (courtesy A. Patnaik).
{\bf b)}\ A 15 GHz radio image of the gravitational lens PKS1830-211.}
\label{mollens}
\end{figure}

\begin{figure}
\psfig{figure=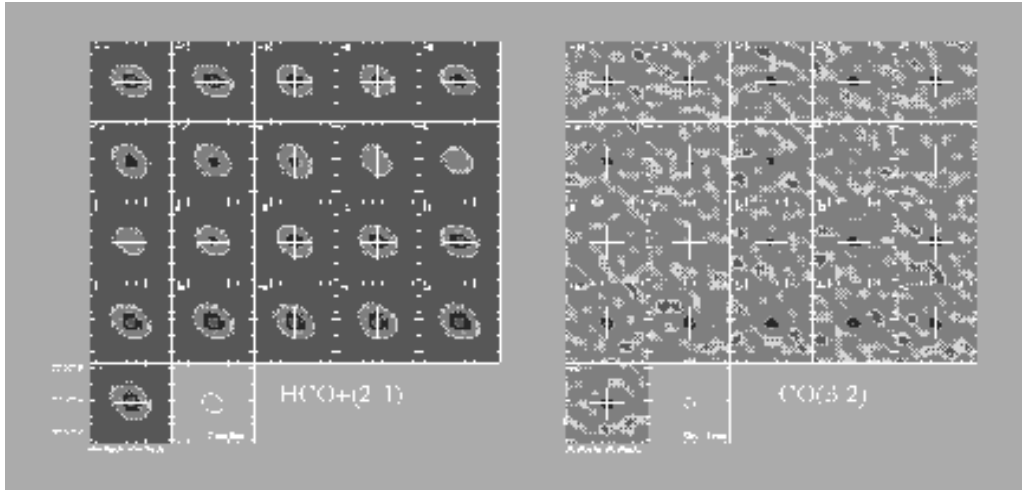,width=13.5cm,angle=0.0}
\caption[]{Channel maps of HCO$^+$(2-1) and CO(3-2) absorption
towards the B0218+357 obtained with the IRAM Plateau de Bure
interferometer. The angular resolution does not resolve the
two lensed images of the background QSO. The continuum weakens
in the channel maps which corresponds to the absorption line,
but never disappears completely. Since the absorption lines are
strongly saturated, this shows that only part of the continuum
is obscured by optically thick molecular gas. From Combes \& Wiklind
(unpublished).}
\label{b0218fig2}
\end{figure}     

\subsubsection{Absorption in gravitational lenses.}\ 
The two absorption line systems with the highest column densities occur
in galaxies which are truly intervening and each acts as a gravitational
lens to the background source: B0218+357 and PKS1830-211. In these
two systems several isotopic species are detected as well as the main
isotopic molecules, showing that the main lines are saturated and
optically thick \cite{combes95} \cite{combes96} \cite{wiklind96a}
\cite{wiklind97}. Nevertheless, the absorption lines do not reach the
zero level. This can be explained by the continuum source being
only partially covered by obscuring molecular gas, but that the obscured
regions are covered by optically thick gas.
The lensed images of B0218+357 and PKS1830-211 consist of two main
components. By comparing the depths of the saturated lines with fluxes
of the individual lensed components, as derived from long radio wavelength
interferometer observations, the obscuration is found to cover only one
of two main lensed components \cite{wiklind95a} \cite{wiklind96a}. This
has subsequently been verified through mm-wave interferometer data
\cite{menten96} \cite{wiklind97} \cite{swift01}.

\subsubsection{B0218+357}\ 
This is a flat-spectrum radio source lensed by an intervening galaxy. The
lens nature was first identified by Patnaik et al.~\cite{patnaik93}. The
lens system consists of two components (A and B), separated by 335
milliarcseconds (Fig~\ref{mollens}a).
There is also a faint steep-spectrum radio ring, approximately centered on
the B component. Absorption of neutral hydrogen has been detected at
$z_{\rm d} = 0.685$ \cite{carilli93}, showing that the lensing galaxy is
gas rich. The redshift of the background radio source is tentatively
determined from absorption lines of Mg\,{\tt II}$\lambda$2798 and H$\gamma$,
giving $z_{\rm s} \approx 0.94$ \cite{browne93}.
Molecular absorption lines were detected in this system \cite{wiklind95a}
further strengthening the suspicion that the lens is gas-rich and likely
to be a spiral galaxy. The molecular absorption lines do not reach zero level.
Nevertheless, absorption of isotopic species show that the main isotopic
transitions must be heavily saturated. In fact, both the $^{13}$CO and
C$^{18}$O transitions were found to be saturated as well, while the
C$^{17}$O transition remained undetected \cite{combes95} \cite{combes97}.
This gives a lower limit to the CO column density which transforms to
$N_{H_2} \approx 4 \times 10^{23}$\,cm$^{-2}$ and an
$A_{\rm V} \approx 850$\,mag.

\begin{figure}
\psfig{figure=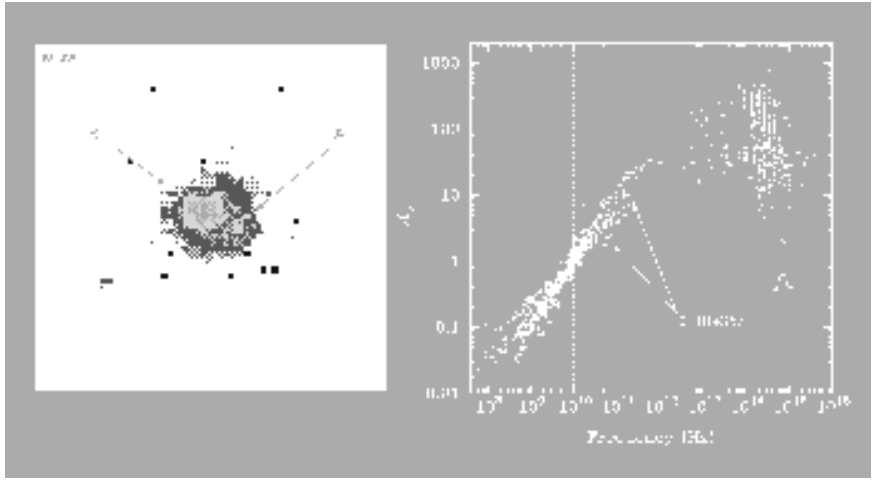,width=11.5cm,angle=0.0}
\caption[]{The gravitational lens B0218+357.
{\bf Left :}\ \ 
Optical image obtained with the HST in the I band (FW814),
showing the A and B components (archival data). In contrast to the
radio image (Fig.~\ref{mollens}a), the A component is weaker than
the B component. This is caused by obscuration of molecular gas, which
gives rise to the observed molecular absorption lines at $z_{\rm d} = 0.688$.
{\bf Right :}\ \ 
Normalized SED for flat-spectrum radio QSOs. The fluxes for the
A component in B0218+357 as observed with the HST are marked by
circles. Their positions suggest that only 3\% of the flux
supposed to come from this component reaches the observer.
(From~\cite{wiklind99}).}
\label{b0218fig3}
\end{figure}

That the molecular gas seen towards B0218+357 covers only one of the
two lensed images of the background source can be seen in
Fig.~\ref{b0218fig2}, where the continuum decreases at velocities
corresponding to the absorption line but never completely disappears.
Subsequent millimeter interferometry observations have shown that the
absorption occurs in front of the A-component, which is then expected
to be completely invisible at optical wavelengths.
Nevertheless, images obtained with the HST WFPC2 in broad V- and I-band,
show both components (Fig.~\ref{b0218fig3}). While the intensity ratio
A/B of the two lensed images is 3.6 at radio wavelengths \cite{patnaik95},
A/B$\approx$0.12 at optical wavelengths. The V$-$I values show no
significant difference in reddening for the A- and B-component.
Hence, there is no indication of excess extinction in front of the
A-component despite the large $A_{\rm V}$ inferred from the molecular
absorption.
Since it is unlikely that the A/B intensity ratio is very much different
at optical and radio wavelengths (differential magnification could
introduce a small difference if the radio and optical emission comes
from separate regions) the A component appears sub-luminous in the
optical. The other possibility is that the B component is over-luminous
at optical wavelengths by a factor 30 (or 1.4 magnitudes), possibly
caused by microlensing.
This latter explanation is, however, quite unlikely in view of the presence
of large amounts of obscuring molecular gas in front of the A component.
By compiling a sample of flat-spectrum radio sources from the literature,
with properties similar to that of B0218+357 (except the gravitational
lensing aspect), correcting for different redshift and normalizing
the observed luminosities at $\nu = 10$ GHz, it is possible to show
that the optical luminosity of B0218+357 is abnormally weak \cite{wiklind99}
(Fig.~\ref{b0218fig3}). In this comparison the observed magnitude
of the A component was used, multiplied by a factor 1.3 in order to
compensate for the B component using the magnification ratio of 3.6.
This clearly showed the A component to be sub-luminous, rather than
the B component being over-luminous.
The interpretation of this is that the A component is obscured by
molecular gas, with an extinction that is very large. Some light `leaks'
out but through a line of sight which contains very little obscuring gas,
hence not showing much reddening in the V$-$I colors. Assuming that
all the obscuration occurs in the A component, only $\sim$3\% of the
photons expected from the A component reaches the observer.
Since the extent of the optical emission region is very small, this
suggests the presence of very small scale structure with a large density
contrast in the molecular ISM of the lensing galaxy.

\subsubsection{PKS1830-211}\ 
This is a radio source consisting of a flat-spectrum radio core and
a steep-spectrum jet. It is gravitationally lensed by a galaxy at
$z_{\rm d} = 0.886$ \cite{wiklind96a} into two images of the
core-jet morphology (Fig.~\ref{mollens}b). The two cores are separated
by 0\ffas97 and the images of the jet form an elliptical ring.
PKS1830-211 is situated close to the Galactic center and suffers
considerable local extinction. Its lens nature was first suspected
through radio interferometry \cite{rao88}, but as neither redshift
was known nor optical identification achieved (cf.~\cite{djorgovski92})
its status as a gravitational lens remained unconfirmed. 

The lensing galaxy was found through the detection of several molecular
absorption lines at $z_{\rm d} = 0.886$ \cite{wiklind96a}.
At millimeter wavelengths the flux from the steep-spectrum jets is
completely negligible and it is only the cores that contributes to the
continuum. It was soon found that the molecular absorption was seen only 
towards one of the cores, the SW image. However, weak molecular absorption
was subsequently found also towards the NE image. This fortunate situation
gives two sight lines through the lens and gives velocity information which
can be used in the lens modeling (see Sect.~\ref{pks1830model}).
A second absorption line system has been found towards PKS1830-211, seen
as 21cm HI absorption at $z=0.19$ \cite{lovell96}, making this
a possible compound lens system. This intervening system complicates the
lens models of this system. A potential candidate for the $z=0.19$ absorption
has been found in HST NICMOS images \cite{lehar00}. It is situated
$\sim$4$^{\prime\prime}$ SW of PKS1830-211 and is designated as G2.
The molecular absorption lines towards PKS1830-211 and their use
for deriving the differential time delay between the two cores will
be described in more detail in Sect~\ref{timedelay} and Sect.~\ref{pks1830model}.

\section{DUST CONTINUUM EMISSION} \label{dustcont}

The spectral shape of the far-infrared background suggests that approximately
half of the energy ever emitted by stars and AGNs has been absorbed by dust
grains and then re-radiated at longer wavelengths \cite{puget96} \cite{fixsen98}
\cite{lagache99} \cite{gispert00}. The dust is heated to temperatures of 20-50
K and radiates as a modified black-body at far-infrared wavelengths.
At the Rayleigh-Jeans part of the dust SED the observed continuum flux increases
with redshift. This is known as a `negative K-correction' and is effective until
the peak of the dust SED is shifted beyond the observed wavelength range, which
occurs at $z>10$. Dust continuum emission from high redshift objects is therfore
observable at millimeter and submillimeter wavelengths and is an important sources
of information about galaxy formation and evolution in general and for
gravitational lenses in particular.

\subsection{Dust emission} \label{dustemission}

Dust grains come in two basic varieties, carbon based and silicon based. Their
size distribution ranges from tens of microns down to tens of {\AA}ngstr\"{o}ms.
The latter are known as PAH's (Polycyclic Aromatic Hydrocarbonates). Except for the
smallest grains, the dust is in approximate thermodynamical equilibrium with
the ambient interstellar radiation field. The dust grains absorb the photon energy
mainly in the UV and re-radiate this energy at infrared and far-infrared (FIR)
wavelengths. The equivalent temperature of the dust grains amount to
15-100 K and they emit as an approximate blackbody.

The spectral energy distribution (SED) of dust emission is usually
represented by a modified blackbody curve,
$F_{\nu} \propto \nu^{\beta} B_{\nu}(T_{\rm d})$ (cf.~\cite{thronson86}
\cite{wiklind95b}), where $B_{\nu}$ is the blackbody
emission, $T_{\rm d}$ the dust temperature and $\nu^{\beta}$ is the
frequency dependence of the grain emissivity, which is in the range
$\beta = 1 - 2$.
Such representations have successfully been used for cold dust components
where a large part of the SED is optically thin. When $\tau \approx 1$ or
larger, the observed dust emission needs to be described by the expression:
\begin{eqnarray} \label{firsedeq}
F_{\nu} & = & \Omega_{s} B_{\nu}(T_{\rm d}) \left(1-e^{-\tau_{\nu}}\right)\ ,
\end{eqnarray}
where $\Omega_{s}$ is the solid angle of the source emissivity distribution,
$\tau_{\nu}$ is the opacity of the dust.
Setting $\tau_{\nu} = \left(\nu/\nu_0\right)^{\beta}$ gives
$F_{\nu} \propto \nu^{\beta}B_{\nu}(T_{\rm d})$ for $\tau_{\nu} \ll 1$
and $F_{\nu} \propto B_{\nu}(T_{\rm d})$ for $\tau_{\nu} \gg 1$.
The critical frequency $\nu_0$ is the frequency where $\tau_{\nu}=1$.

\subsubsection{The infrared luminosity.}\ 
The total infrared luminosity is derived by integrating Eq.~\ref{firsedeq}
over all frequencies.
Here the flux density $F_{\nu}$ corresponds to the energy emitted by dust only.
The infrared luminosity for an object at a redshift $z$ is given by
\begin{eqnarray} \label{firlum}
L_{\rm IR} & = & 4\pi \left(1+z\right)^{3} D_{\rm A}^{2}
\int\limits^\infty_0{F_{\nu} d\nu}\ ,
\end{eqnarray}
where $D_{\rm A}$ is the angular size distance\footnote{Expressing Eq.~\ref{firlum}
in a form directly accessible for integration, we get
\begin{eqnarray} \label{firint}
{{L_{\rm IR}} \over {{\rm L_{\odot}}}} & = & 
8.53 \times 10^{10}\,\left(1+z\right)^{3}\,\left[{{D_{\rm A}} \over {{\rm Mpc}}}\right]^{2}
T_{\rm d}^{4} \Omega
\int\limits^\infty_0{{{x^3 \left(1 - e^{-\left(a x\right)^{\beta}}\right)} \over
{e^{x}-1}}dx}\ .\nonumber
\end{eqnarray}
The integral can be integrated numerically with appropriate values of
the parameter $a = kT_{\rm d}/h\nu_0$.}.
The solid angle $\Omega$ appearing in Eq.~\ref{firsedeq} is a parameter
derived in the fitting procedure. In the event of a single dust component,
$\Omega$ can be estimated from the measured flux $F_{\nu_{\rm r}}$ at
a given restframe frequency $\nu_r$
\begin{eqnarray} \label{omega}
\Omega & = & {{F_{\nu_r}} \over
{B_{\nu_r}(T_{\rm d})\left(1-e^{-\left(\nu_r/\nu_0\right)^{\beta}}\right)}}
\nonumber \\
 & \approx & 
6.782 \times 10^{-4}\left[{{\nu_r} \over {{\rm GHz}}}\right]^{-3}
\left[{{{\rm F_{\nu_r}}} \over {{\rm Jy}}}\right]
\left[{{e^{h\nu_r/kT_{\rm d}}-1} \over 
{1-e^{-\left(\nu_r/\nu_0\right)^{\beta}}}}\right]\ .
\end{eqnarray}
Using some typical values ($T_{\rm d} = 30$ K, $\beta = 1.5$,
$\nu_0 = 6$ THz ($50\mu$m), $\nu_{\rm r} = \nu_{\rm obs}(1+z) =
1400$ GHz ($\nu_{\rm obs} = 350$ GHz at $z=3$) and, finally,
an observed flux of 1\,mJy) we get $\Omega \approx 2 \times 10^{-14}$.
For a spherical source with a radius $r \approx D_{\rm A}\sqrt{\Omega/\pi}$,
this corresponds to a dust continuum emission region with an extent of
only $\sim 160$\,pc.

Although this is a very rough estimate of the size of the emitting region,
it shows, since typical observed values were used, that FIR dust emission
from distant objects tend to come from very small regions. This will be
of importance when considering the effects of gravitational lensing.

\begin{figure}
\psfig{figure=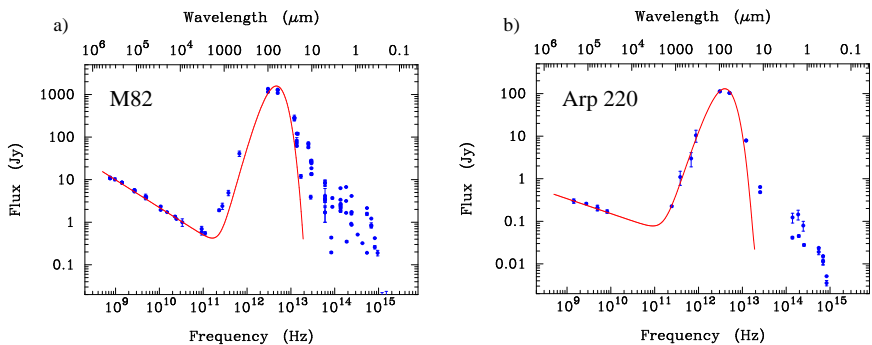,width=11.5cm,angle=-90.0}
\caption[]{The spectral energy distribution of two starburst galaxies.
{\bf a)} M82 and {\bf b)} Arp220. Despite a difference in far-infrared
luminosity of almost 2 orders of magnitude, their spectral energy distribution
are nearly identical. Notice also the presence of cold dust in M82, visible
as an excess flux at millimeter and submillimeter wavelengths.
The SEDs have been fitted by a modified blackbody curve, which becomes optically
thick at 50$\mu$m and which has $\beta = 2.0$ for M82 and $\beta = 1.3$ for Arp220
and using a single temperature component of T$_{\rm d} = 45$ K for both galaxies.
}
\label{firseds}
\end{figure}

\subsubsection{The dust mass.}\ 
An estimate of the dust mass from the infrared flux requires either
optically thin emission combined with a knowledge of the grain properties,
or optically thick emission and a knowledge of the geometry of the
emission region (cf.~\cite{hildebrand83}).

The grain properties are characterized through the macroscopic mass
absorption coefficient, $\kappa_{\nu}$.  Several attempts to estimate
the absolute value of $\kappa_{\nu}$ as well as its frequency dependence
have given different values (cf.~\cite{hughes97}).
Combining the same frequency dependence as in \cite{hughes97} with
a value given by \cite{hildebrand83}, the mass absorption coefficient
can be described as
\begin{eqnarray}
\kappa_{\nu_r} & \approx 0.15 & \left({{\nu_r} \over
{375{\rm GHz}}}\right)^{1.5}\ {\rm m^{2}\ kg^{-1}}\ ,
\end{eqnarray}
where $\nu_{\rm r}$ corresponds to the restframe frequency.
This expression corresponds to a grain composition similar to that
found in the Milky Way. At frequencies where the emission is
optically thin, the dust mass can now be determined from\footnote{
Eq.~\ref{mdust} can also be expressed as:
\begin{eqnarray}
M_{\rm d} & \approx &
4.08 \times 10^{4} \times \nonumber
\left[{{F_{\nu_{obs}}} \over {{\rm Jy}}}\right]
\left[{{D_{\rm A}} \over {{\rm Mpc}}}\right]^{2}
\left({{\nu_r} \over {375{\rm GHz}}}\right)^{-9/2}
\left(e^{h\nu_r/kT_{\rm d}}-1\right)(1+z)^{3}\ \ 
{\rm M_{\odot}}\ .\nonumber
\end{eqnarray}
},
\begin{eqnarray} \label{mdust}
M_{\rm d} & = &
{{F_{\nu_{obs}}} \over {\kappa_{\nu_r} B_{\nu_r}(T_{\rm d})}}
D_{\rm A}^{2}(1+z)^{3}\ .
\end{eqnarray}

\subsection{Detectability of dust emission} \label{dustdetect}

A typical far-infrared spectral energy distribution (SED) of a starburst
galaxy (M82) is shown in Fig.~\ref{firseds}a. The SED of a more powerful
starburst (Arp220) is shown in Fig.~\ref{firseds}b. Perhaps the most striking
aspect of these SEDs is their similarity, despite that they represent
galaxies with widely different bolometric luminosities. In both cases most
of the bolometric luminosities comes out in the far-infrared: M82 has a
far-infrared luminosity of $3 \times 10^{10}$\,L$_{\odot}$, while Arp220 is
a so called  Ultra-Luminous Infrared Galaxy (ULIRG) with a far-infrared
luminosity of $1 \times 10^{12}$\,L$_{\odot}$.
The SEDs shown in Fig.~\ref{firseds} have been fitted by a modified
blackbody curve, which becomes optically thick at 50$\mu$m and which has
$\beta = 2.0$ for M82 and $\beta = 1.3$ for Arp220 (cf. Eq.~\ref{firsedeq}).
The modified blackbody curves has been fitted using a single temperature
component of T$_{\rm d} = 45$ K for both galaxies. Notice, however, the
presence of a colder dust component in the SED of M82, which is visible
as an excess flux at millimeter and submillimeter wavelengths \cite{thuma00}.

The observed dust continuum emission originates from dust grains in
different environments and which are heated by different sources.
Nevertheless, a remarkably large number of dust SEDs, like the ones
shown in Fig.~\ref{firseds}, can be well fitted by only one, or in some
cases two dust components (cf. the cold dust component in M82).

\medskip

For a single dust temperature component, the flux ratio between the
submillimeter (850$\mu$m) and the far-infrared (100$\mu$m) is strongly
dependent on the dust temperature. For T$_{\rm d} = 45$ K (as in the
case of M82 and Arp220), $f_{850\mu{\rm m}}/f_{100\mu{\rm m}}
\approx 3 \times 10^{-3}$, while for T$_{\rm d} = 20$ K, 
$f_{850\mu{\rm m}}/f_{100\mu{\rm m}} \approx 0.06$, or 20 times
larger. Nevertheless, as long as the dust temperature is not extremely low,
it is much harder to observe the long wavelength tail of the dust SED than
the peak at $\sim$100$\mu$m (except that in the latter case one
needs to observe from a satellite due to our absorbing atmosphere).

\begin{figure}
\psfig{figure=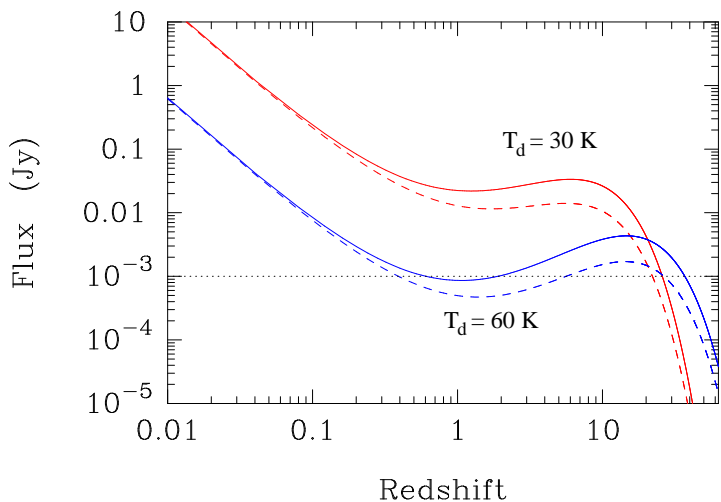,width=9.5cm,angle=-90.0}
\caption[]{The observed flux density at 850$\mu$m of a source with
$L_{\rm FIR} = 5 \times 10^{12}$\,L$_{\odot}$ as a function of redshift.
The top set of curves correspond to a dust temperature of 30 K and the bottom
curves to a dust temperature of 60 K. The full drawn lines correspond
to a flat matter dominated universe ($\Omega_{\rm m} = 1.0, \Omega_{\Lambda} = 0$),
and the dashed curve to a flat $\Lambda$ dominated universe ($\Omega_{\rm m} = 0.3,
\Omega_{\Lambda} = 0.7$).}
\label{firsed}
\end{figure}

At millimeter and submillimeter wavelengths the SED can, to a first approximation, be
characterized by $f_{\nu} \propto \nu^{\gamma}$, where $\gamma = 3-4$.
Hence, the observed flux increases as an object is shifted to higher
redshift. This effect is large enough to completely counteract the
effect of distance dimming.
An example of this is shown in Fig.~\ref{firsed}, where the observed flux at 850$\mu$m
has been calculated for a FIR luminous, $5 \times 10^{12}$ L$_{\odot}$, galaxy,
for two different dust temperatures, T$_{\rm d} = 30$ K and T$_{\rm d} = 60$ K,
and for two different cosmologies.
The largest uncertainty in the predicted flux as a function
of redshift comes from the assumed dust temperature, rather than the
assumed cosmology. However, regardless of dust temperature and cosmology,
the effect of the `negative K-correction' of the dust SED is to make
the observed flux more or less constant between redshifts of $z=1$
and all the way to $z\approx10$, where the Wiener part of the modified
blackbody curve is shifted into the submillimeter window and the flux
drops dramatically.

This constant flux over almost a decade of redshift range makes the
millimeter and submillimeter window extremely valuable for studies of
the formation and evolution of the galaxy population at high redshift
in general and for gravitational lensing in particular. For a constant
co-moving volume density, the submm is strongly biased towards detection
of the highest redshift objects. The prerequisite is, of course, that
galaxies containing dust exist at these large distances and that the low
flux levels expected can be reached by our instruments.
Both of these criteria are actually fulfilled; powerful new bolometer
arrays working at millimeter (MAMBO, and recently SIMBA) and in the
submillimeter (SCUBA) have shown that low flux levels can be observed
and that objects containing large amounts of dust do exist at early
epochs (cf.~\cite{hughes98} \cite{smail98b} \cite{eales99}).

\subsection{Submillimeter source counts} \label{numbercounts}

One of the first studies using long wavelength radio continuum emission
was to simply count the cumulative number of detected sources as a function
of flux level. These observations mainly probed high luminosity radio
galaxies and showed a significant departure from an Euclidean non-evolving
population. This was the first evidence of cosmic evolution \cite{ryle67}
\cite{jauncey75}.

The negative K-correction in the mm-to-far-infrared wavelength regime for
dust emission has enabled present day submm/mm telescopes, equipped with
state-of-the-art bolometer arrays, to get a first estimate of the source
counts of FIR luminous sources at high redshift.
There are two bolometer arrays which have produced interesting results
so far; SCUBA on the JCMT in Hawaii and MAMBO on the IRAM 30m telescope
in Spain. Additional arrays are under commissioning and will likely 
contribute to this area shortly: SIMBA on the 15m SEST on La Silla, and
BOLOCAM on the 10m CSO on Mauna Kea.

\begin{figure}
\psfig{figure=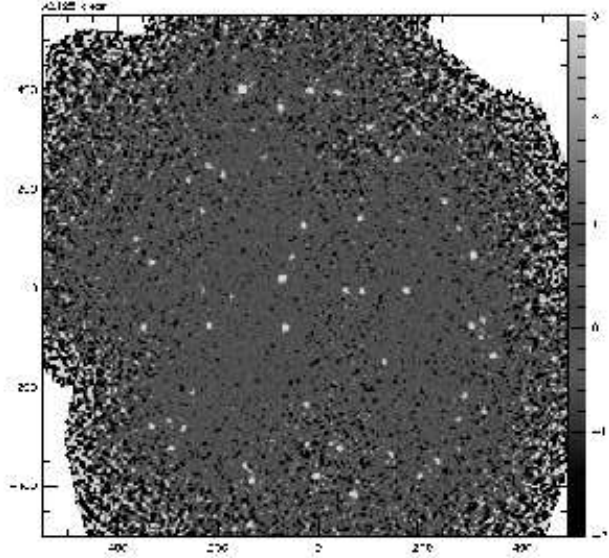,width=11.0cm,angle=-90.0}
\caption[]{An image of the cluster Abell 2125 obtained with the MAMBO bolometer
array at the IRAM 30m telescope (from~\cite{carilli00b}). The angular size
is in arcseconds. The noise rms is at 0.5 mJy/beam.}
\label{mambo}
\end{figure}

The SCUBA bolometer array at the JCMT was put to an ingenious use when
it looked at blank areas of the sky chosen to be towards rich galaxy clusters
at intermediate redshift \cite{smail97} \cite{smail98b} \cite{barger99}
\cite{blain99b} \cite{blain99c}. The gravitational magnification by the
cluster enabled very low flux levels to be reached and several detections
were reported. This method has been used by others as well and an example
of an image of the rich cluster Abell 2125 at 1250$\mu$m is shown in
Fig.~\ref{mambo} \cite{carilli00b}. More than a dozen sources are detected
above the noise but none is associated with the cluster itself. Instead
they are all background sources gravitationally magnified by the cluster
potential.

\medskip

The cumulative source count of a population of galaxies is simply the surface
density of galaxies brighter than a given flux density limit. In a blank field
observation it is in principle derived by dividing the number of sources with
the surveyed area. The effects of clustering has to be considered if the observed
area is small. In practice there are several statistical properties that have
to be considered. Usually the threshold for source detection is not uniform
across the mapped area. Since the sources are generally found close to the
detector limit those which have fluxes boosted by spurious noise has a
higher likelihood to be detected than those which experience a negative
noise addition, which are likely to be lost from the statistics. This latter
effect leads to an overestimate of the true source flux. The possibility of
an additional bias through differential magnification will be discussed in
Sect.~\ref{diffmag}.

The case of submm/mm detected galaxies behind foreground galaxy clusters is
yet more complicated (cf.~\cite{blain99b}).  The gravitational lens distorts
the background area and magnifies the source fluxes. The magnitude of these
effects may vary across the observed field. A detailed mass model of the lens
is needed in order to transform the observed number counts into
real ones, as well as knowledge about the redshift distribution of the sources.
Smail and collaborators (\cite{smail97} \cite{smail98a}\cite{smail98b})
initially observed 7 clusters, constructed or used existing mass models of
the cluster potentials, and managed to obtain source counts at sub-mJy
levels (cf.~\cite{blain99b}). Although the lensing effect of clusters
allows observations of weaker fluxes, it introduces an extra uncertainty
in the number counts. This is, however, not dominating the overall error
budget \cite{blain99b}. There is another beneficial effect with the
lensing in that the extension of the background area alleviates the problem
of source confusion. The angular resolution of existing bolometer arrays
is approximately 15$^{\prime\prime}$ and source confusion is believed
to be a problem at flux levels below 0.5 mJy.

Other blank field surveys using SCUBA have pushed as deep as the cluster surveys,
but without the extra magnification they probe somewhat higher flux levels.
Examples of such deep blank field surveys include the Hubble Deep Field North
\cite{hughes98}, the fields used for the Canada-France Redshift Survey
\cite{eales99} \cite{eales00}, the Lockman hole and the Hawaii deep field region
SSA13 \cite{barger98}.

All these submm deep fields, including the cluster fields, are only a few
square arcminutes. Using on-the-fly mapping techniques a few groups have recently
started mapping larger areas but to a shallower depth (cf.~\cite{borys01}
\cite{scott01}).

\begin{figure}
\psfig{figure=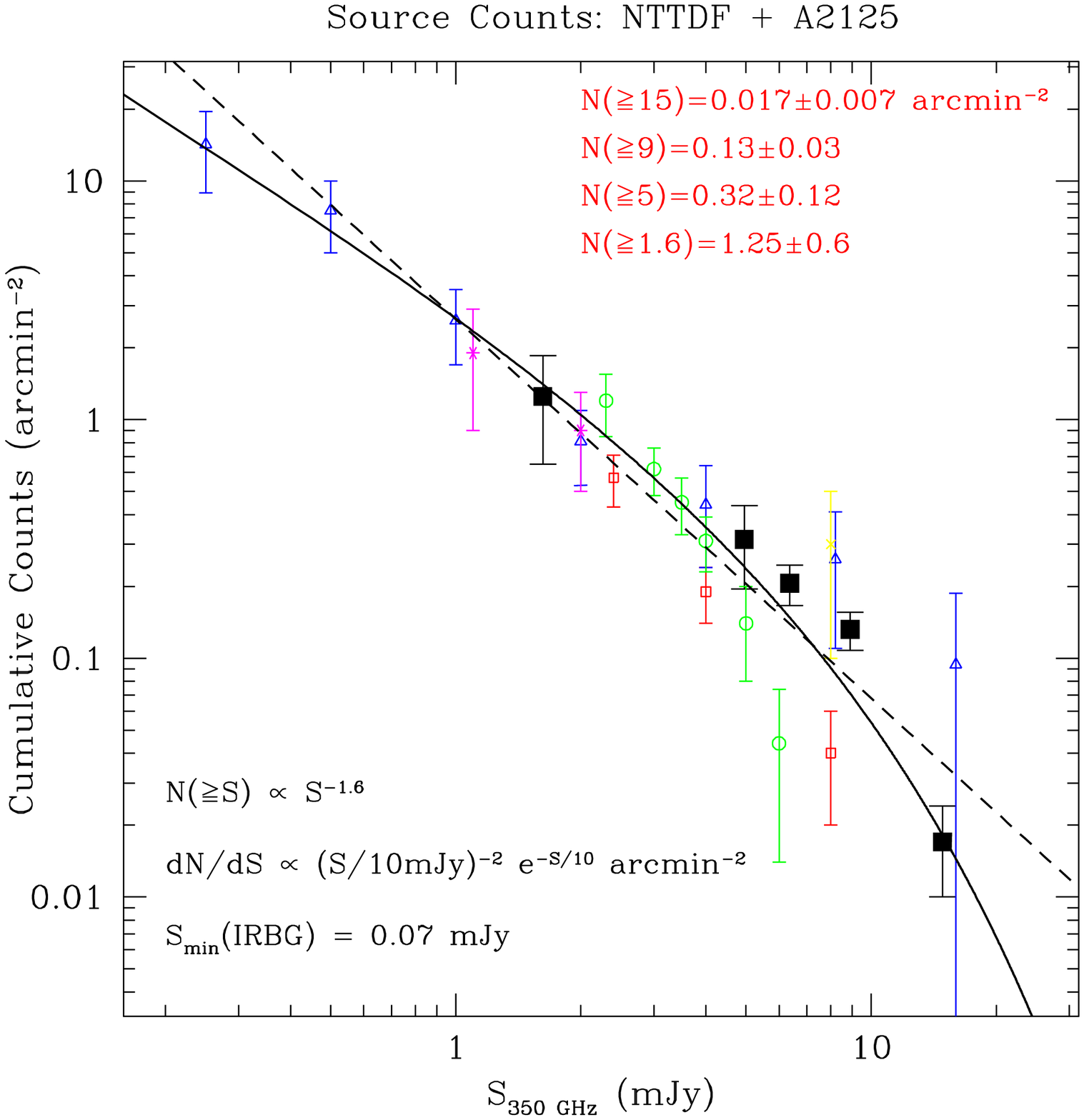,width=11.5cm,angle=0.0}
\caption[]{Source counts from several surveys using SCUBA at 850$\mu$m and
MAMBO at 1250$\mu$m (from~\cite{carilli00b}). The dashed curve is a powerlaw
of index $-1.8$ while the solid curve is an integrated Schechter luminosity
function with a powerlaw index $-2$ and an exponential cut-off at 10mJy.
All fluxes refer to 850$\mu$. The 1250$\mu$m data points have been multiplied
by a factor 2.25 in order to transform them into expected fluxes at 850$\mu$m.}
\label{counts}
\end{figure}

Carilli et al.~\cite{carilli00b} combined the number counts from all the
blank-field observations. The result is a cumulative source count stretching
from $\sim$15 mJy to 0.25 mJy (Fig.~\ref{counts}). The source counts obtained
using the lensing technique, after correcting for the lensing effects, are
compatible with those obtained through pure blank-fields.
The turnover at a flux level of $\sim$10 mJy is probably real and represents a
maximum luminosity of $\sim 10^{13}$ L$_{\odot}$ for an object at $z \approx 3$.
The exact shape of the number counts is still uncertain at both the low and
high flux ends. Results from the MAMBO bolometer array, which operates at 1250$\mu$m,
have been multiplied by a factor 2.25 in order to transform it into the expected
flux at 850$\mu$m. This assumes that the objects have an SED of the same type as
starburst galaxies (cf. Fig.~\ref{firseds}).

\medskip

In order to transform the cumulative source count into a volume density
it is necessary to know the redshift distribution of the sources. It is,
however, possible to circumvent this by fitting a model of galaxy evolution
to the observed source counts. This has been explored extensively by Blain
et al.~\cite{blain99c}, (see also~\cite{combes99a} \cite{takeuchi01}), and
will not be discussed further here.

\subsection{Submm source identification and redshift distribution}

The sources detected in submm/mm surveys can in a majority of cases be
identified with sub-mJy radio sources (cf.~\cite{smail00}). This population
of weak radio continuum sources is believed to be powered by star formation
rather than AGN activity \cite{windhorst85} \cite{haarsma00}. Attempts
to identify the submm/mm sources with optical and/or infrared counterparts have
failed in all but a small number of cases (cf.~\cite{downes99a} \cite{ivison00}
\cite{frayer00}).
The submm/mm detected population is not related to nearby nor intermediate
redshift sources, but are believed to be at $z > 1$, but the lack of clear
optical/IR identifications has made it difficult to assess its true redshift
distribution. An alternative technique for determining the redshift has been
introduced by Carilli \& Yun~\cite{carilli00a}, which relates the radio
continuum flux at 1.4 GHz with the measured flux at 850$\mu$m (see
also~\cite{barger00}). As the radio flux declines with increasing redshift,
the submm flux increases (cf. Fig.~\ref{firseds}). Although the method is
model dependent (mainly depending on the dust temperature T$_{\rm dust}$,
the radio spectral index as well as the frequency dependence of the dust
emissivity coefficient, cf. Sect.~\ref{dustemission}), it gives a rough
estimate of the redshift. Using this method it has been possible to show
that the majority of the submm/mm detected sources lie at a redshift
$1 \leq z \leq 4$ (cf.~\cite{smail00} \cite{carilli00b}).

\subsection{Differential magnification} \label{diffmag}

One well-known property of gravitational lensing is that it is
achromatic, meaning that the deflection of photons by a gravitational
potential is independent of wavelength. The achromaticity is applicable
to observed gravitational lenses as long as the source size is small
compared to the caustic structure of the lens, such as when the
Broad Line Region (BLR) of a QSO is lensed by a galaxy sized lens.
Chromatic effects can, however, become important if the source is
substantially extended (relative to the caustic structure) and the
spectral energy density of the source is position dependent.

The submm/mm detected dusty sources discussed in Sect.~\ref{dustcont}
are characterized by extended emission, several orders of magnitude
larger than the compact sources generally studied in gravitational
lensing. This applies to dust emission regardless whether the dust is
heated by star formation or by a central AGN. Measured on galactic scales,
however, the dust is relatively centrally concentrated, with typical
scales ranging from $10^2$ pc to a few kpc (cf. Sect.~\ref{dustemission}).
A dust distribution heated by a central AGN will have a radial dust
temperature distribution, even when radiation transfer effects and a
disk- or torus-like geometry are considered. This is observed in nearby
Seyfert galaxies \cite{polletta99}.
A radial temperature profile is also found in the case of a pure starburst
\cite{siebenmorgen99}, but spatially more extended than in the AGN case.
Gravitational lensing of an extended dust distribution with a non-homogeneous
temperature, and thus emissivity distribution, means that the assumption of
achromaticity is no longer valid and the source may be differentially
magnified.

\medskip

If the characteristic length scale in the source plane is $\eta_0$ the
characteristic length scale in the lens plane is
$\xi_0 = ({\rm D}_{\rm d}/{\rm D}_{\rm s})\,\eta_0$. Taking a dust distribution
of 1 kpc ($\eta_0$), a source redshift $z_{\rm s} = 3$ and a lens redshift
$z_{\rm d} = 1$, the characteristic length scale in the lens plane becomes
approximately 0\ffas2. This is close to the typical image separation for
strong lensing. Since the submm/mm detected galaxies are believed to show a
significant change in the dust temperature over this scale, it is quite likely
that they will exhibit chromatic effects.

\medskip

An analytical model of the effect of differential magnification of dusty
sources was presented in \cite{blain99a}, where it was shown that the effect
can be strong and that it was most likely to produce an increase in the
mid-infrared flux relative to the long wavelength flux. This would make
the sources appear warmer than what their intrinsic SED would imply.

\begin{figure}
\psfig{figure=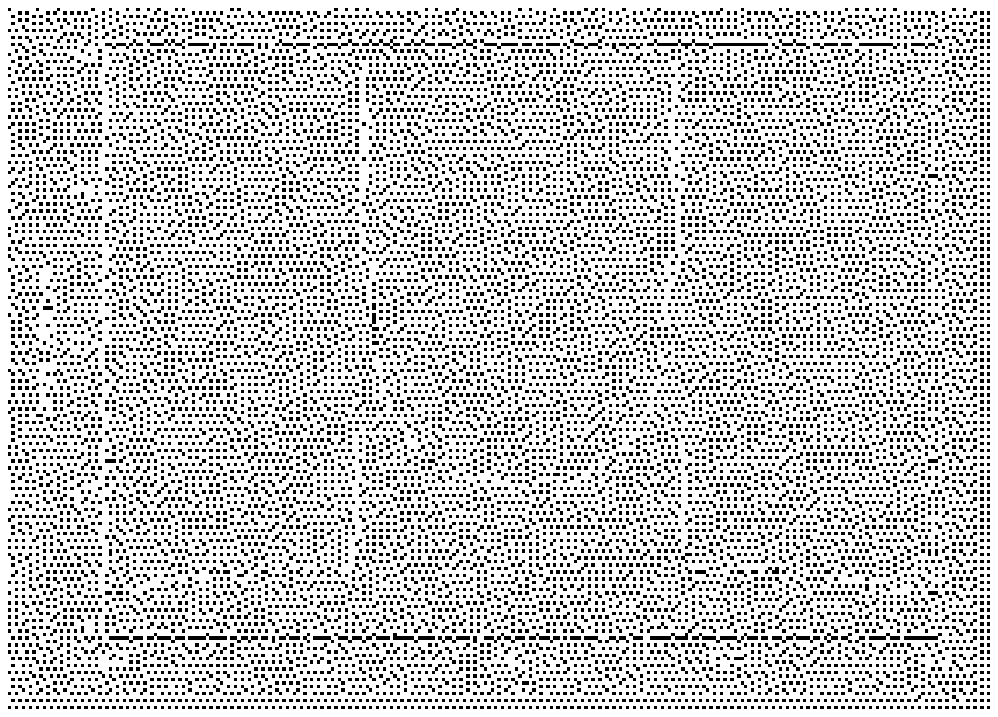,width=11.5cm,angle=-90.0}
\caption[]{Magnification in a cut through the caustic structure of an elliptical
lens configuration. The two peaks corresponds to the radial caustic. This particular
cut does not pass through the tangential caustic, which would have produced a yet
stronger magnification peak close to the center. The length scale is in
arcseconds. Notice the very strong gradient in the magnification when going
from the one-image region to the the extended three-image `plateau'. The
magnification changes by a factor $\sim 10$ over angular scales of $\sim 0\ffas01$,
corresponding to scales of $\sim 50$ pc at $z_{\rm s} \approx 2$.
From K. Pontoppidan's Master Thesis, Copenhagen University \cite{pontoppidan00}.}
\label{diffmag3}
\end{figure}

A more detailed analysis of the effect of differential magnification and
its probability for occurance was done by Pontoppidan \cite{pontoppidan00}.
Using elliptical potentials and a realistic parameterization of the dust
and its spectral energy distribution it was showed that both positive and
negative distortions of the SED can occur. Here positive means an increase
in the mid-IR part and negative means an increase in the far-IR/submm part.
Fig.~\ref{diffmag3} shows a plot of the magnification in a cut through one
of these models (elliptical potential) which does not hit the inner tangential
caustic. By placing the center of a dust emission region, with a radial
temperature profile, well outside the radial caustic (i.e. $>|0\ffas5|$
from the center), parts of the outer region of the dust distribution will
fall on the high magnification plateau inside the radial caustic and be
multiply imaged while the center is singly imaged and only moderately
magnified. This situation would cause an enhancement of the long wavelength
part of the SED relative to the mid-IR part. The radius of a typical
dust distribution is $\sim 0\ffas2$ at $z \approx 3$. If the center of the
source is placed closer to the radial caustic, both the center and the extended
dust distribution will be magnified, but the warmer central dust will experience
a larger average magnification and hence result in a flattening of the SED at
mid-IR wavelengths. Again, it might be that the cool dust is multiply imaged
while the center (possibly containing an AGN) is singly imaged.

Pontoppidan \cite{pontoppidan00} found that the cross section for an
enhancement of the long wavelength part of the SED is larger than for
an enhancement of the mid-IR. However, the latter situation results in a
stronger magnification and effects the observed SED to a higher degree.
The latter case also represents a situation where the system is more
likely to be recognized as a gravitational lens.

\begin{figure}
\psfig{figure=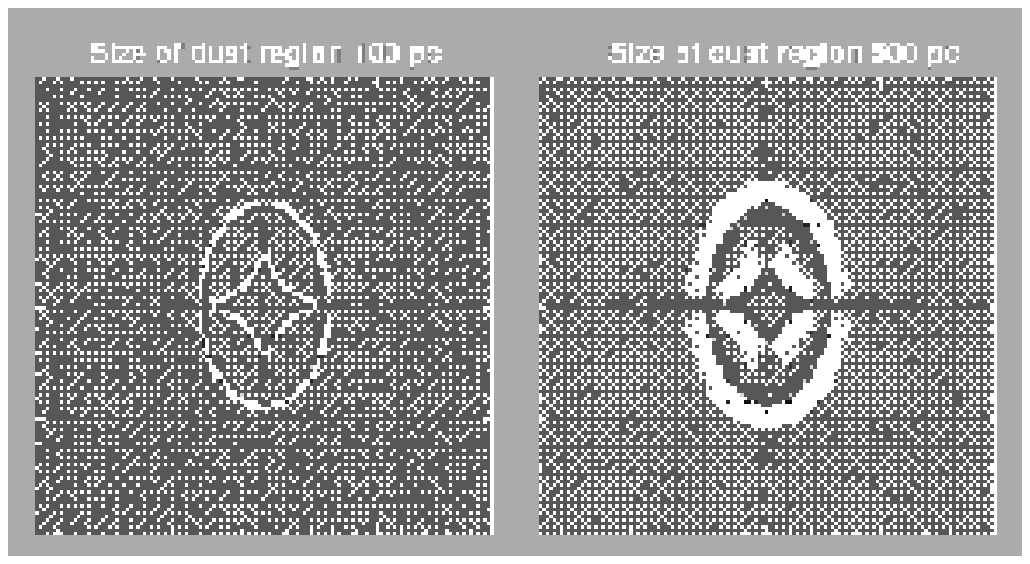,width=11.5cm,angle=0.0}
\caption[]{The caustic structure of an elliptical potential showing areas
where the dust spectral energy distribution will be influenced by differential
magnification. The dust emission region is modeled as a circular disk with a
radial temperature profile. The central heating can be either an AGN or a
dense starburst. If the center of the dust emission region is placed in the
black area, the observed SED will appear cooler than the intrinsic one, while
the opposite effect occurs if the center is located in the black areas.
The effect of a small (left) and large (right) dust region is illustrated.
From K. Pontoppidan's Master Thesis, Copenhagen University \cite{pontoppidan00}.}
\label{diffmag2}
\end{figure}

\subsubsection{Effect on number counts of submm/mm detected galaxies.}\ 
A distortion map of the effect of differential magnification is shown in
Fig.~\ref{diffmag2} (from~\cite{pontoppidan00}). The caustic structure of an
elliptical potential representing a central mass surface density of
$3 \times 10^{9}$ M$_{\odot}$\,kpc$^{-2}$, where the distortion of the
dust SED due to differential magnification has been color coded. Black
represents a negative distortion (cooler SED) and white represents a positive
distortion (warmer SED). The dust distribution of the source is assumed
to have a radius of 100 pc (left image) and 500 pc (right image). Quite
naturally, the larger the region over which dust is distributed, the larger
is the region where negative distortion can occur. By placing the center
of the source in the black/white regions of the distortion map, the observed
SED will appear cooler/warmer.

The implications for the submm/mm detected objects at high redshift is that
differential magnification could induce a bias in the number counts. Especially
since in most of the surveys done so far the sources are found close to
the detection limit. The effect could induce an overestimate of the number
of sources but it could also influence the slope of the cumulative number
counts. The latter is more likely but better statistics from surveys reaching
low noise levels are needed, as well as a better understanding of the
total cross section for positive/negative distortions of dusty submm/mm
sources.

An interesting consequence of the differential magnification of these sources
is that some, perhaps several, of the detected submm/mm objects may be
multiply imaged systems when viewed at high angular resolution in submm/mm
wavelengths. The radio identifications that have been done typically reach
a resolution of 1$^{\prime\prime}$ which is not sufficient to see multiple
images on the expected 0\ffas1-0\ffas2 scale. High angular resolution deep
imaging with future instruments such as ALMA will resolve this issue.

\section{CASE STUDIES} \label{casestudies}

In order to describe in more detail the characteristics of millimeter observations
and interpretations of gravitationally lensed sources, as well as to illustrate
their use, three cases are presented below. First is
the luminous Broad-Absorption-Line (BAL) quasar APM08279+5255 at $z=3.9$.
The gravitational lens hypothesis for this source was put forward based
only on its apparent luminosity. The second case is a detailed study of the
quadruply lensed Cloverleaf quasar, where the gravitational lensing of molecular
gas has enabled a more detailed and constrained lens model.
The last example is PKS1830-211, where the lensing galaxy was actually first
detected through millimetric molecular absorption lines at $z_{\rm d} = 0.886$.
The molecular absorption lines in this system has been used to constrain the
lens model by giving the velocity dispersion and are used to derive the differential
time delay between the two main lensed components.

\subsection{APM08279+5255: A case of differential magnification?} \label{apm08279}

This object was discovered serendipitously during a search for Galactic
carbon stars \cite{irwin98}. It was found to be a BAL QSO at a redshift
$z=3.911$ (see~\cite{downes99b} for the redshift determination). With an
astounding R-band magnitude of 15.2 and detection in three of the four IRAS
bands, its bolometric luminosity turns out to be $5 \times 10^{15}$ L$_{\odot}$.
This in itself led to the suspicion that it is a gravitationally lensed object
Subsequent observations, both from the ground and from space
\cite{ledoux98} \cite{egami00} \cite{ibata99}, led to the detection
of three components, with a maximum separation of $0\ffas35 \pm 0\ffas02$, and
with a flux ratio of the two brightest components of $1.21 \pm 0.25$
(cf. Fig.~\ref{apmfig3}). The optical spectra of the two main components are
similar to each other \cite{ledoux98}.
No lensing galaxy has been identified, although the weak third image could
potentially be the lens (see below). Nevertheless, based on the small separation
of the main components, their similar spectra and the enormous luminosity inferred
for the system, the lensing nature of this system is not questioned. Even in the
case of strong gravitational magnification, APM08279+5255 is an intrinsically very
luminous system, with L$_{\rm bol} \geq 10^{13}$ L$_{\odot}$.

The high apparent brightness of APM08279+5255 has allowed a very good S/N optical spectra
of the intervening absorption line systems to be obtained with the HIRES spectrograph
on Keck \cite{ellison99}. Several potential lens candidates are found as
Mg{\tt II} absorption line systems, with the most conspicuous one at $z=1.181$.
Placing the third image at this redshift, however, requires the lens to be unusually compact
and luminous. It would need to be almost 5 magnitudes brighter than an L$^{\star}$
galaxy with the relevant velocity dispersion of $\sim 150$ km\,s$^{-1}$ \cite{ibata99}
\cite{egami00}. The possibility that the lens harbors an AGN can be dismissed
since no emission lines from $z < 3.9$ are detected in the spectrum. Also, the continuum
of an intervening QSO should have been detected in the saturated parts of the absorption
lines seen towards the background source. No such emission is detected
(cf.~\cite{ellison99}).
APM08279+5255 could thus represent a `text book' example of a gravitational lens
with an odd number of components.

\medskip

\begin{figure}
\psfig{figure=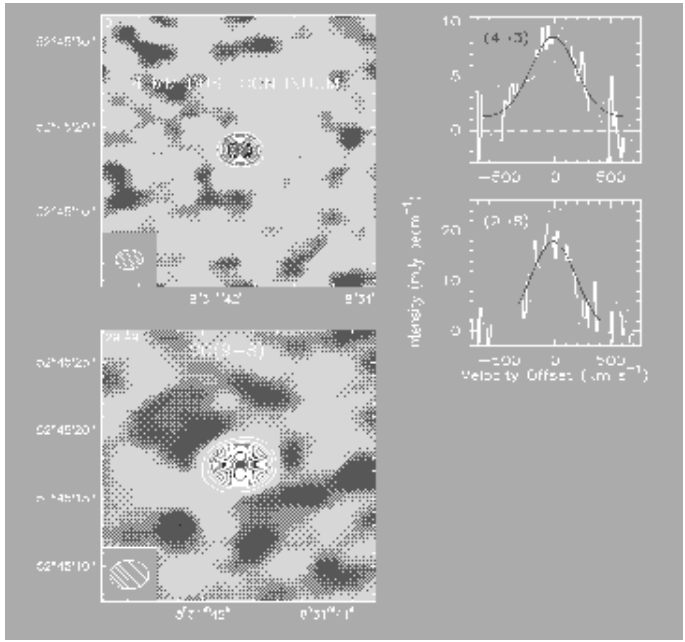,width=9.0cm,angle=0.0}
\caption[]{CO spectra and maps of APM08279+5255 observed with the IRAM Plateau de
Bure interferometer (Downes et al.~\cite{downes99b}). {\bf Upper left:}\ 1.4mm dust continuum.
{\bf Lower left:}\ CO(9-8) emission. {\bf Upper and lower right:}\ CO(4-3) and CO(9-8)
emission line profiles. The angular resolution of the maps is $3\ffas2 \times 2\ffas3$,
far too coarse to resolve the individual components seen at optical/NIR wavelengths.
The maps are centered on $08^{\rm h}31^{\rm m}41\ffs70,\ +52^{\circ}45^{\prime}17\ffas35$
(J2000).}
\label{apmfig1}
\end{figure}

\begin{figure}
\psfig{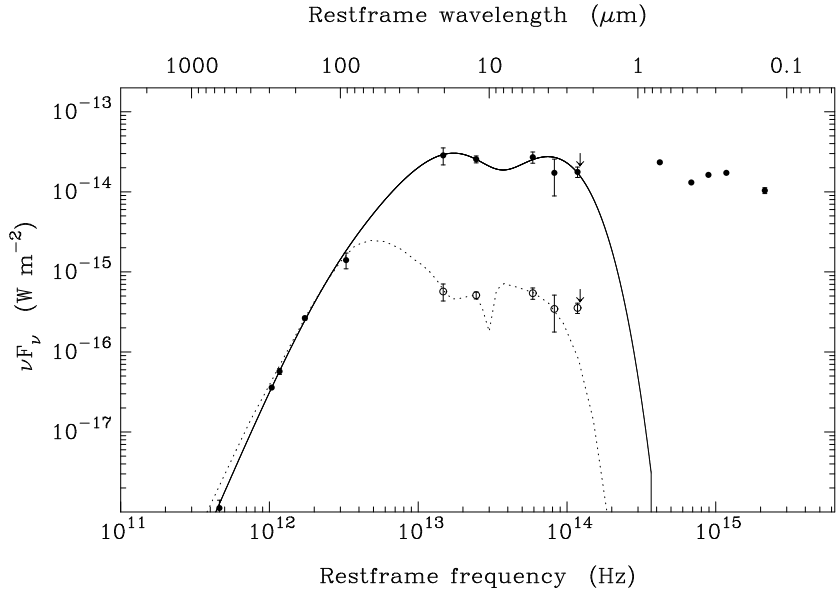}
\caption[]{The spectral energy distribution of APM08279+5255 fitted by two isothermal
greybody models. The full-drawn line corresponds to the sum of a `cool' component
(T$_{\rm dust} = 200$ K) and a hot component (T$_{\rm dust} = 910$ K). The dust
emission becomes optically thick at $\lambda < 200\mu$m.
The data points (filled circles) are from Irwin et al.~\cite{irwin98}, Ledoux et al.~\cite{ledoux98},
Lewis et al.~\cite{lewis98}, Downes et al.~\cite{downes99b}, Egami et al.~\cite{egami00}
and the Faint Source IRAS catalog (NED).
The dotted line is a starburst model of Rowan-Robinson \& Efstathiou \cite{rowanrobinson93}, which has
been arbitrarily fitted to the long wavelength part of the SED. The data points marked
by open circles are the equivalent observed data points but with their values reduced
by a factor 50 in order to fit on the starburst model.}
\label{apmfig2}
\end{figure}

\begin{figure}
\psfig{figure=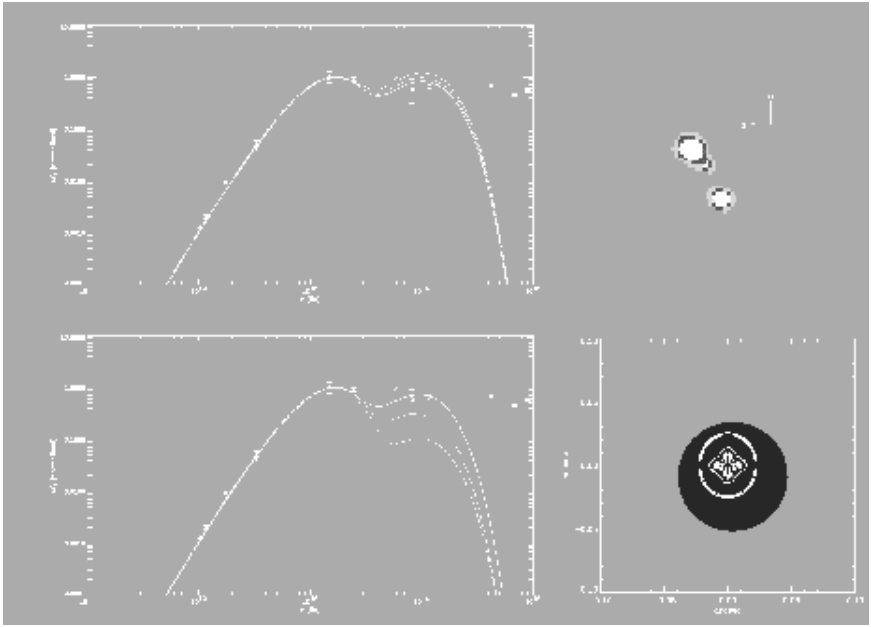,width=11.5cm,angle=0.0}
\caption[]{The 2-image and 3-image lens model solutions for APM0927+5255 and
their effect on the spectral energy distribution. In the 2-image solution (top
left), the intrinsic mid-IR part of the SED is slightly depressed, while in the 3-image
solution (bottom left) the intrinsic mid-IR is strongly enhanced (representing a
positive distortion, see Sect.~\ref{diffmag}).
The intrinsic SED is represented by dashed and dotted lines, for a dust distribution
with a radius of 600 pc and 300 pc, respectively. The full drawn line is the observed
SED.
In the top right is APM08279+5255 seen with the HST WFPC2 camera and
lower right is the dust model and its location used in the 3-image solution.
From K. Pontoppidan's Master Thesis, Copenhagen University \cite{pontoppidan00}.}
\label{apmfig3}
\end{figure}

Apart from being luminous at optical and UV wavelengths, APM08279+5255 also contains
large amounts of dust and metal rich molecular gas (Fig.~\ref{apmfig1}).
The SED of APM08279+5255 is actually dominated by a strong dust continuum emission
(Fig.~\ref{apmfig2}), detected over a wide wavelength band: from the restframe submm
to mid-infrared bands. This puts APM08279+5255 in the class of hyperluminous IR galaxies
even when correcting for a strong gravitational magnification.

The overall dust spectral energy distribution is characterized by a steeply
rising long wavelength part, with a change of slope around
$\lambda_{\rm rest}=200\mu$m, and a flat mid-IR part.
The dust continuum spectra can be fitted by two dust components. One `cool'
characterized by a dust temperature of T$_{\rm d} = 200$ K, which is optically
thin at $\lambda > 200\mu$m (cf. Fig.~\ref{apmfig2}). The second component
is hot, with T$_{\rm d} \approx 910$ K, close to the sublimation temperature of
carbon based dust grains. This second component is optically thick.  The total
dust mass, uncorrected for gravitational magnification, is
$2 \times 10^{8}$ M$_{\odot}$, most of it contained in the cool dust component.

The CO emission lines shown in Fig.~\ref{apmfig1} includes the high excitation
transition $J=9-8$. The CO $J=9$ level is $J(J+1) \times 2.77 = 249$ K above the
ground state. Normal type Galactic molecular clouds with typical H$_2$ densities
of $\sim 300 - 10^3$ cm$^{-3}$ are not sufficient to collisionally populate the
CO $J=9$ level. The mere detection of the CO(9-8) line therefore shows that the
gas has to be unusually dense and warm. This immediately suggests that this
gas component resides close to the QSO, possibly associated with the hot dust
component. If both the CO $J=4-3$ and $J=9-8$ emission are associated with the
same gas component, the total molecular gas mass, corrected for magnification,
is quite modest: $3 \times 10^9$ M$_{\odot}$ \cite{downes99b}. If the
lower transition, on
the other hand, emanates from a more extended and cooler region than the $J=9-8$
transition, the total molecular gas mass can be one to two orders of magnitude
larger. That this is likely to be the case was shown by the detection of CO $J=1-0$
and $J=2-1$ emission from APM08279+5255 (Papadopolous et al. 2001)\footnote{The
offset between the CO emission presented in Papadopoulos et al.~\cite{papadopoulos01} and that
of Downes et al.~\cite{downes99b} results from the use of slightly different coordinates
for APM08279+5255. The coordinates given in the caption of Fig.~\ref{apmfig1}
corresponds to the best optical/IR coordinates determined from both ground and
space based imaging.}. Using the same conversion factor between H$_2$ column
density and velocity integrated CO intensity as is used for the Milky Way and
nearby galaxies, the total molecular gas mass in APM08279+5255, uncorrected
for gravitational magnification, is $(0.6-3.2) \times 10^{11}$ M$_{\odot}$
\cite{papadopoulos01}. The amount of gravitational magnification is in
this case expected to be low due to the extended nature of the molecular gas,
especially the gas seen in the lower transitions.
Incidentally, three additional CO emitting sources are detected within
3$^{\prime\prime}$ of the center of APM08279+5255 \cite{papadopoulos01}.
If these are not gravitationally lensed images, which they are not if the currently
best lens models are used (cf.~\cite{ibata99} \cite{egami00}), these three
additional sources are not magnified to any significant degree. The field around
APM08279+5255 should then represent a remarkable over-density of gas rich galaxies
at high redshift. These three additional sources are, however, not detected in
continuum emission with the Plateau de Bure interferometer nor at optical or
NIR wavelengths and their exact nature remains undetermined.

\medskip

APM08279+5255 has a SED which is essentially flat from a restframe wavelength of
$\sim$30$\mu$m to optical wavelengths (cf. Fig.\ref{apmfig2}). This is usually
interpreted as being the effect of a face-on configuration of a dust-disk surrounding
a central AGN. The low inclination of the disk enables the observer to get an
un-obscured view of the hot dust close to the AGN as well as the cool dust further
away. Comparison with dust models calculated by Granato et al. \cite{granato96}
\cite{granato97} shows that the mid-IR slope is too shallow even for the most extreme
face-on models \cite{lewis98}, i.e. the dust SED in APM08279+5255 appears to be too
`warm' even if heated by a powerful AGN.
Another possible explanation for the flat mid-IR SED is that APM08279+5255 experiences
differential magnification of the dust emission region. This possibility was explored
by Egami et al.~\cite{egami00} by applying sources of various sizes to their lens model.
In Fig.~\ref{apmfig2} the SED of APM08279+5255 is shown together with a starburst
model \cite{rowanrobinson93}. The starburst model has been arbitrarily fitted to the
long wavelength part of the observed SED. At mid-IR wavelengths, the starburst model
predicts a flux which is $\sim$50 times lower than the observed fluxes in APM08279+5255.
The data points marked by open circles are the equivalent observed fluxes diminished
by a factor of 50 in order to fit the starburst model. Although APM08279+5255
undeniably contains a powerful AGN, which is likely to contribute a substantial
part of the heating of the gas and dust, the influence of star formation can not be
ruled out. Can differential magnification account for at least part of the difference
between a pure starburst SED and the observed one?

\subsubsection{Modeling the lens APM08279+5255.}\ 
The lensing configuration of this system has been modeled by Egami et
al.~\cite{egami00} and Ibata et al.~\cite{ibata99}. Using an isothermal
elliptical potential with no external shear, two different types of lens
models were applied: a three-image model and a two-image model. The
former model is non-singular in order to produce the third image, while
the latter assumes that the third image is the lensing galaxy and the potential
is singular in order to suppress the formation of the third image.
The two-image model produce a modest magnification of $\sim$7
(cf.~\cite{egami00}), while the three-image model produce a magnification
of $\sim$90 for both a point source and a more extended source distribution
\cite{ibata99} \cite{egami00}. Since the apparent bolometric luminosity
exceeds $10^{15}$ L$_{\odot}$, the three-image configuration is more appealing.
However, the core radius is large, 0\ffas21, almost as large as the Einstein
radius, 0\ffas29. If the lens is at a redshift $z_{\rm d} \approx 1.2$,
the core radius corresponds to $\sim$1.2 kpc in the lens. This is much
larger than most measured core radii. In a survey of 42 giant elliptical
galaxies it was found that the objects which can be resolved have a median
core radius of 225$h^{-1}$ pc \cite{lauer85}. In the case of a three-image
model, the potential is almost circular with $\epsilon = 0.012$, while
the two-image model gives $\epsilon = 0.083$ \cite{egami00}. The difference
in ellipticity corresponds to a difference in the size of the caustic
structure, which in turn influences the effects of differential magnification.
In the three-image model, the caustic structure is approximately 45 pc in extent
in the source plane, while the two-image model has a caustic structure almost
5 times larger.

\medskip

The effects on the lensing behavior for a source with a finite extent was explored
in \cite{egami00}. A source with an extent exceeding $\sim500$ pc resulted in a
filled disk. A more detailed model was done by Pontoppidan \cite{pontoppidan00}
where an assumed source temperature distribution was used in order to derive
the resulting spectral energy distribution. Using the model parameters of
\cite{egami00}, where the source is located between the radial and tangential
caustic for the three-image model, the hot dust is expected to be moderately
enhanced by the outer magnification plateau (cf. Fig.~\ref{diffmag2}). In
the two-image scenario, the QSO is again located outside the tangential caustic.
In this case, however, the radial caustic is lacking due to the singular
potential. The latter scenario can produce a modest negative distortion of
the SED (as seen in the top left panel of Fig.~\ref{apmfig3}). In the
three-image scenario, however, the effect on the SED is more dramatic and
represents a positive distortion, i.e. the mid-IR part of the SED is enhanced
relative to the long wavelength part (bottom left panel of Fig.~\ref{apmfig3}).
The magnitude of the distortion is quite large, its details depending on the
extent of the dust region. For a dust distribution with a radius of 650 pc, the
differential magnification can enhance the intrinsic flux at restframe mid-IR
wavelength with a factor $\sim10$. A smaller extent of the dust results in a
smaller enhancement factor. The dust region (for a radius of 300 pc) and the
caustic structure are seen in the lower right panel of Fig.~\ref{apmfig3}.

The three-image model of APM08279+5255 is more likely to be correct than the
two-image model since it produces a magnification which corresponds to a source
with a bolometric luminosity which is large, but not extreme.
The three-image model also means that for realistic dust distribution, the
restframe mid-IR is strongly enhanced relative to the longer wavelength part
of the SED. The intrinsic SED of APM08279+5255 resembles that of less extreme
dusty QSO spectra (cf.~\cite{carilli00b}). In fact, the shape of the
intrinsic SED of APM08279+5255 now resembles that of pure starburst models,
except that the mid-IR is still enhanced by a factor $5-10$, marking the
influence of the AGN. This shows that the effects of differential magnification
must be considered before applying radiation transfer models to gravitationally
lensed dusty sources.

In order to model the resolved and extended low-J CO emission, the effects of
a highly elliptical lensing potential has been explored \cite{lewis02}.
The lensing galaxy is here assumed to be an edge-on spiral. The result is
a good fit with the extended low-excitation CO emission, while the point sources
from the background QSO, although imaged into three components, have widely
different magnification ratios compared to the observed values. This may not
be of great importance if microlensing affects the optical photometric results
(e.g.~\cite{lewis99}).

\subsection{The Cloverleaf: Another case of differential magnification} \label{cloverleaf}

The Cloverleaf is the gravitationally lensed image of the BAL quasar
H1413+117 at $z=2.558$, showing four quasar-images (hereafter called spots)
with angular separation from 0\ffas77 to 1\ffas36. Since its discovery
\cite{magain88}, the Cloverleaf has been imaged with ground based
telescopes in numerous bands up to I and with HST/WFPC2 in the UV, optical
and near-IR \cite{turnshek97} \cite{kneib98a} \cite{kneib98b}.
   
\subsubsection{The lensing system.}\ 
After the early lens model of Kayser et al.~\cite{kayser90}, these new data sets 
have been used to derive an improved model of the lensing system \cite{kneib98a}
\cite{kneib98b} which now includes:

\begin{enumerate}

\item A cluster of galaxies with derived photometric redshifts in the 
range 0.8 to 1.0, which contributes to the magnification.

\item A lensing galaxy close to the line of sight to the quasar, which 
determines the geometry of the image (four main spots) and carries
the largest share of the magnification. The redshift of the 
lensing galaxy has been tentatively measured with VLT/ISAAC at a 
value of 0.9 (Faure et al. in prep).

\end{enumerate}

In the following, we use this new model for the lensing system, which 
is essentially constrained by the HST data. Further details can be 
found in \cite{kneib98a} \cite{kneib98b}.

\subsubsection{The IRAM millimeter data sets.}\ 
After its discovery in the CO(3-2) line emission with the IRAM Pico 
Veleta dish \cite{barvainis94}, the Cloverleaf has been observed 
in the millimeter range by various teams and instruments (\cite{wilner95}
with BIMA, \cite{yun97} with OVRO). Yet, the best data sets
collected to date on this object are from the IRAM Pico Veleta dish 
and Plateau de Bure interferometer.

A total of six millimeter transitions have been reported from 
observations with the IRAM Pico Veleta dish: CO(3-2), CO(4-3), CO(5-4), 
CO(7-6), CI($^3$P$_1$-$^3$P$_0$) and HCN(4-3) \cite{barvainis97}.
Detailed non-LTE modeling of the CO line strengths by these authors
indicates that the molecular gas is warm (T larger than 100 K), dense
( n(H$_2$) density larger than $3 \times 10^3$ cm$^{-3}$) and not very
optically thick ($\tau_{\rm CO} < 3$). These results suggest that the
molecular material is close to a powerful heating source and might
therefore be related to the environment of the central engine in the
quasar. They also prompt us for not using the conventional conversion
factor CO to H$_2$ which is derived for molecular clouds in the disk
of our Galaxy.

\begin{figure}
\psfig{figure=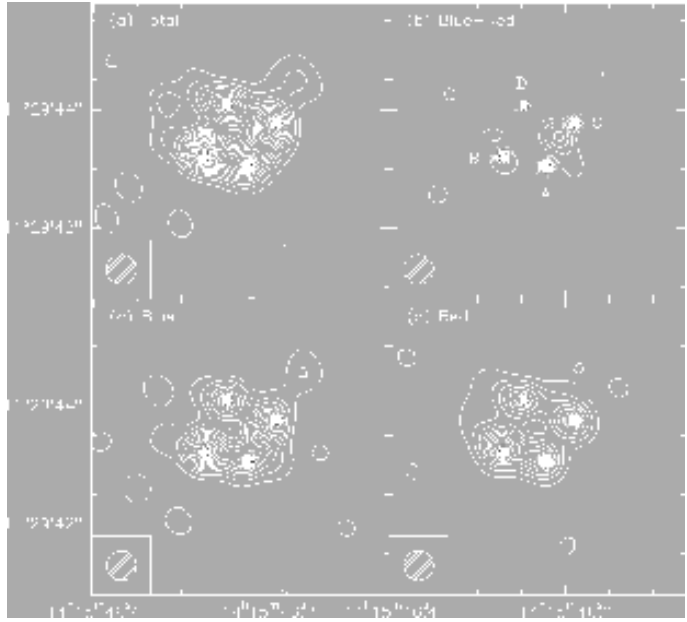,width=9.0cm,angle=0.0}
\caption[]{Image of the Cloverleaf obtained with the IRAM Plateau de Bure
interferometer \cite{kneib98a}. {\bf a)} is the total CLEANed image.
{\bf c)} and {\bf d)} are the blue and red part of the total emission
profile, while {\bf b)} shows the difference between the CLEANed blue and
red images. A velocity gradient in the underlying source can be inferred
from the residual. The data has been restored with a circular beam of angular
resolution 0\ffas5.}
\label{cloverfig1}
\end{figure}

\begin{figure}
\psfig{figure=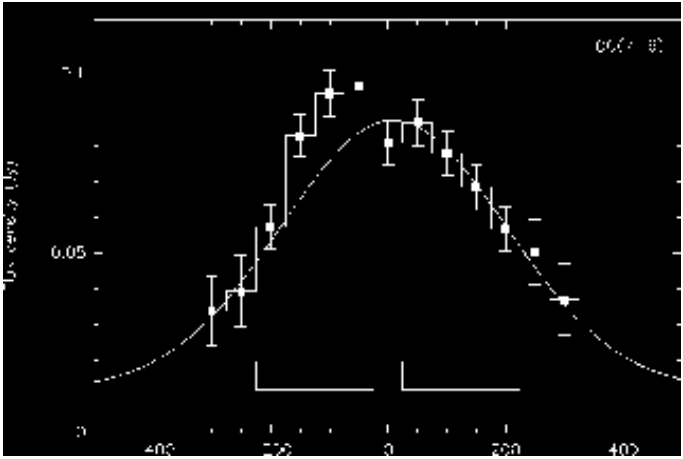,width=9.0cm,angle=0.0}
\caption[]{The $^{12}$CO(7-6) spectrum from the $z=2.56$ Cloverleaf quasar
obtained with the IRAM Plateau de Bure interferometer \cite{kneib98a}.
The thin line represents a best fit Gaussian profile. Notice the asymmetric
line profile with excess emission at the blue part of the spectrum.}
\label{cloverfig2}
\end{figure}

Thanks to the strength of the CO(7-6) transition, a high resolution
(0\ffas5) map was obtained with the IRAM Plateau de Bure interferometer.
A first CO(7-6) interferometric data set \cite{alloin97} has later
been complemented with observations at intermediate baselines \cite{kneib98a}. 
The combined data has lead to the CLEANed map restored 
with an 0\ffas5 circular beam, shown in Fig.~\ref{cloverfig1} .
In order to search for a velocity gradient, we have derived the spatially
integrated line profile, following the procedure described in~\cite{alloin97}.
The CO(7-6) line profile (Fig.~\ref{cloverfig2}) shows a marked
asymmetry with a steep rise and excess of emission (with respect to a
standard Gaussian) on its blue side and a slower decrease on its red side.
Excluding the central velocity channel (so that the split in
velocity is symmetric), we have built the blue (-225, -25 km/s) 
and the red (+25, +225 km/s) maps displayed in Fig.~\ref{cloverfig1}c,d 
respectively. The difference between the red  and blue CLEANed 
maps (Fig.~\ref{cloverfig1}b) establishes firmly the presence of a velocity
gradient at the 8$\sigma$ level. Measurements of the characteristics
of the spots from the CO(7-6) image have been performed (spot flux 
ratios, sizes and orientations) through a fitting procedure in the 
visibility domain, as explained in~\cite{alloin97}. Final 
parameters are provided in Table 1 of \cite{kneib98a} where the 
spot sizes are intrinsic to the image, i.e. deconvolved by the
interferometer beam: spots A, B and C definitely appear elongated. 

\subsubsection{Comparing images in the UV and the millimeter range.}\ 
Images in the UV/optical correspond in the quasar restframe to
the emission from the accretion disk surrounding the quasar central
engine. This latter source is expected to be point-like. The four spots 
on the HST images do indeed have a stellar-like appearance, being
circular with a FWHM of about 0\ffas068 \cite{kneib98a}.
The absolute photometry and relative intensity ratios of the four
spots (Table 4 in \cite{kneib98a}) have been computed using the
Sextractor software \cite{bertin96}. The large variation
of the intensity ratios in U, compared to V, R and I bands can
probably be explained by absorption along the line of sight by
intervening galaxies. Alternatively, this effect can be ascribed
to dust extinction at the redshift of the quasar, using an SMC-like
dust extinction law \cite{turnshek97}. The presence of such an
absorbing medium in the close environment of the central engine
could be put in relation with the BAL appearance of this quasar.

The millimeter CO lines are expected to arise from an extended
structure, the so-called dusty/molecular torus, with an intrinsic
radius of a few 10 pc to a few 100 pc (according to models).
In such a configuration, different parts of the extended torus will 
be positioned differently with respect to the caustic (the curve 
which represents in the source plane the signature of the lensing 
system). As the image properties and the amplification factor in 
particular, are ruled by the relative positioning of the source/caustic,
the four spots on the CO(7-6) image, each corresponding to the extended 
torus, will be distorted with respect to the four spots on the HST 
image (corresponding each to a point-like source). This features what 
is called `differential magnification effects' (see Sect.~\ref{diffmag}).
From the blue bump appearing on the CO(7-6) line profile
(Fig.~\ref{cloverfig2}), we clearly see that the blue-shifted part
of the CO line arises from a region of the  molecular torus which is
positioned closer to the caustic than the region emitting the red-shifted
side of the CO line. In this way, we are able to recover detailed
structural and kinematical information about the molecular torus in
the quasar.

In order to derive precisely the shear induced by the lensing system
on the extended source in the quasar, it is imperative to register 
with a high accuracy the Cloverleaf image in a waveband corresponding 
to a point-like source in the quasar (accretion disk: UV restframe, 
that is an R band image for example) and in a waveband corresponding 
to an extended source in the quasar (molecular torus: CO(7-6) line). 
The high precision required, better than 0\ffas2, was achieved using
a combination of the HST data and of CFHT data acquired over a larger 
field of view under extremely image quality \cite{kneib98a}.

\begin{figure}
\psfig{figure=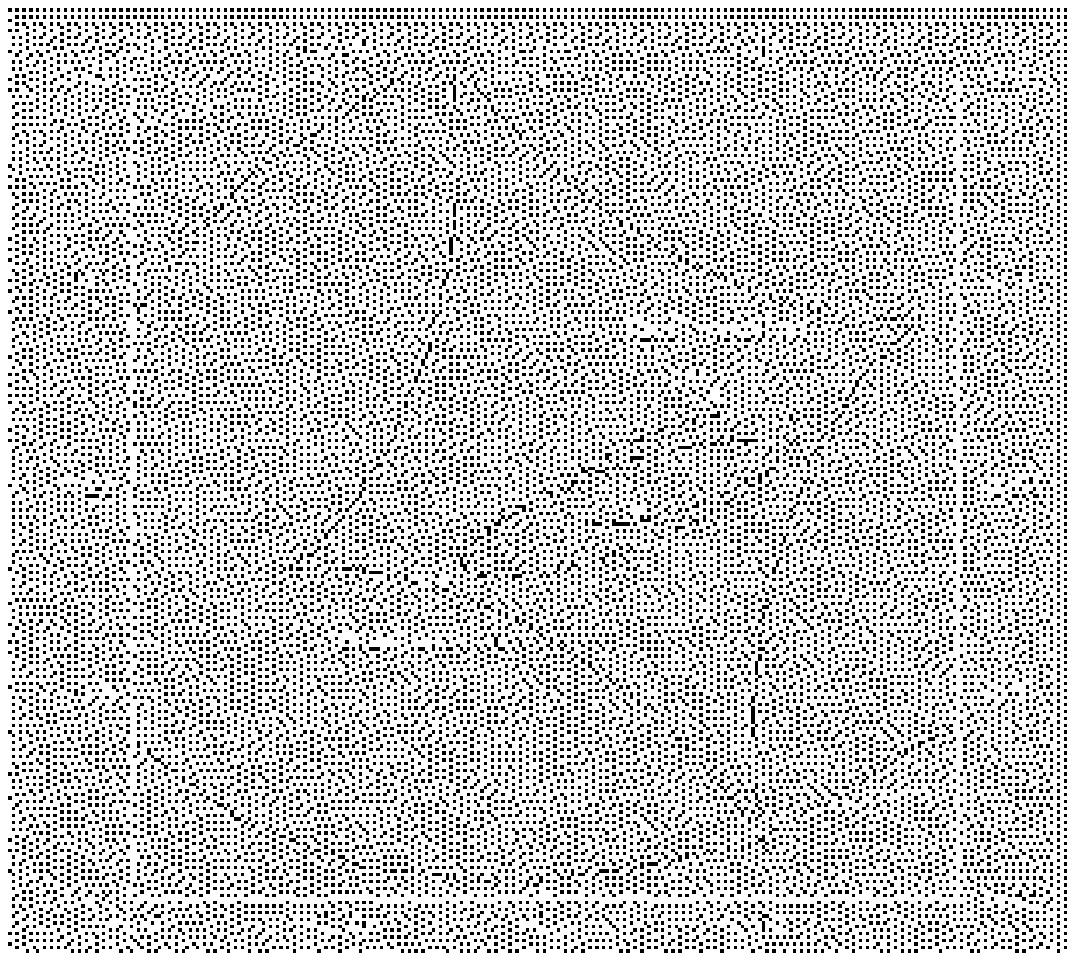,width=8.0cm,angle=0.0}
\caption[]{The caustic structure and the CO source distribution in the
Cloverleaf quasar \cite{kneib98a}. The central ellipses represent
the CO source distribution for the red and blue part of the emission
profile, respectively. The dot in the center corresponds to the quasar
UV point source. The scale on the axis is arcseconds. The result
has been obtained by combining an HST optical image with the interferometric
CO data.}
\label{cloverfig3}
\end{figure}

\subsubsection{Derived properties of the molecular torus in the Cloverleaf BAL
quasar at $z=2.558$.}\ 
We have used the model of the lensing system as constrained by the 
HST data and presented above. The total amplification factor for 
the CO emission is found to be 30. This amplification factor translate
to a molecular gas mass M(H$_2)=2 \times 10^9$ M$_{\odot}$ and an
atomic hydrogen gas mass M(HI)$=2 \times 10^9$ M$_{\odot}$
\cite{barvainis97} \cite{kneib98a}. 

We have derived the properties of the molecular torus in the quasar 
using the CO(7-6) maps: firstly using the total line flux and 
secondly, using separately each of the blue and red halves of the 
CO(7-6) line \cite{kneib98a}. We find a typical size for the 
molecular torus of 150 pc (assuming H$_0=50$ km s$^{-1}$, $\Omega_m=1$ and 
$\Omega_{\Lambda}=0$). When we treat separately the maps corresponding
to the blue-half line and red-half line, we find that the quasar
point-like UV source is almost exactly centered between the region
emitting the blue-half of the CO line and the region emitting the red-half
of the  line (Fig.~\ref{cloverfig3}). This is reminiscent of a disk- or
ring-like structure orbiting the quasar at about 75 pc and with a Keplerian
velocity of 100 km\,s$^{-1}$ (assuming a disk orientation perpendicular
to the plane of the sky). The resulting central dynamical mass would be
about $8 \times 10^8$ M$_{\odot}$. This value is in good agreement with
the estimate of the molecular gas mass made above from the total CO line flux,
provided uncertainties in the inclination of the molecular torus and in the
conversion factor from I(CO) to N(H$_2$).

In conclusion, we can regard the case of the Cloverleaf as a first 
and enlightening example of what will become routine when ALMA becomes
available. Indeed, exploiting differential magnification effects is an
extremely promising technique. The effective angular resolution on the
CO source in the quasar at $z=2.558$, using this procedure, is $\sim$0\ffas03,
or about 17 times smaller than the synthesized beam of the IRAM
observations! And the amplification factor in this case is around 30!

\subsection{PKS1830-211: Time delay and the Hubble constant} \label{timedelay}

Observations and applications of differential time delays in gravitational
lenses are discussed in detail elsewhere in this book.
Here some results which have implications for the millimetric part of the
electromagnetic spectrum will be presented. More specifically, we will 
discuss a derivation of the differential time delay in the gravitational
lens PKS1830-211 obtained from the saturated molecular absorption line
of HCO$^+$(2-1).

The gravitational lens system PKS1830-211 has been described in some
detail above (Sect.~\ref{molabs}).
The background quasar is variable at radio wavelengths, with an amplitude
which increases at shorter wavelengths. This is due to the fact that the
core, where the variability occurs, is a flat-spectrum source while the
jet, which has a more or less constant flux, has a steep radio spectrum.
It is presently unknown if PKS1830-211 is variable at infrared/optical
wavelengths, although this is likely to be the case.

\medskip

Time delay measurements of the PKS1830-211 system has also been done using
long wavelength radio continuum \cite{ommen95} \cite{lovell98}.
In one case a single dish telescope was used and the two main lens components
were not resolved \cite{ommen95}. The analysis had to be based on a compound
light curve and the derived time delay of $44 \pm 9$ days should therefore
be regarded as tentative.
In the other case, the ATCA interferometer was used, with an angular resolution
that did not fully resolve the NE and SW components \cite{lovell98}.
Instead a model fitting procedure was used in order to obtain two separate
light curves over an 18 months period. The resulting differential time
delay is $26^{+4}_{-5}$ days. Although the analysis is model dependent,
this result represents a considerable improvement in the $\Delta t $
estimate.

\subsubsection{Time delay measurements using molecular absorption lines.}\ 
As discussed in Sect.~\ref{molabs}, the lens in the PKS1830-211 system was
first detected through molecular absorption lines at a redshift
$z_{\rm d} = 0.88582$ \cite{wiklind96a}.
More than 16 different molecular species in 29 different transitions
have so far been detected at millimeter wavelengths \cite{wiklind96a}
\cite{wiklind98} \cite{gerin97}. Two additional molecular species
in three different transitions have been observed at cm wavelengths
\cite{menten99}.

\begin{figure}
\psfig{figure=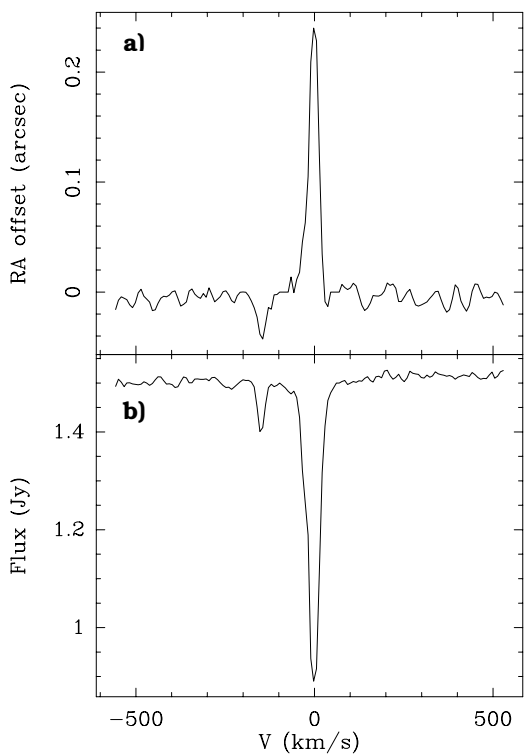,width=8.0cm,angle=0.0}
\caption[]{{\bf Bottom:}\ Spectrum of HCO$^+$(2-1) at
$z=0.886$ towards PKS1830-211 obtained with the IRAM interferometer.
The main absorption line is seen around zero velocity. A secondary,
weaker absorption line of HCO$^+$(2-1) is seen at a velocity of
$-147$ km/s relative to the main line.
{\bf Top :}\ The right ascension shift of the phase center of the
continuum emission as a function of velocity. A negative shift means
that the phase center moves towards the NE component, while a positive
shift indicates a shift towards the SW component.
Comparison with the absorption spectra shows that the main
absorption component covers the SW source, while the weaker secondary
absorption covers the NE source.
(From~\cite{wiklind98}.}
\label{pks1830spec1}
\end{figure}

The millimeter transitions include three different isotopic variants:
H$^{13}$CO$^+$, HC$^{18}$O$^+$ and H$^{13}$CN. The mere detection of
these lines shows that the main isotopic transitions of these 
molecules must be highly saturated. Despite this the absorption lines
do not reach zero intensity (Fig.~\ref{pks1830spec1}b). This
can only be reconciled with an optical thick obscuration that do
not completely cover the background continuum emission. In fact,
from the ratio of the total continuum and the depth of saturated
molecular absorption lines (such as HCO$^+$(2-1), HCN(2-1), etc.),
it was concluded that only the SW lens component is obscured by
molecular gas and that the covering factor of this particular image
is unity or close to unity \cite{wiklind96a}. A secondary weaker
molecular absorption has now been found towards the NE component
as well, separated in velocity by $-147$ km\,s$^{-1}$ \cite{wiklind98}.

Imaging of the HCO$^+$(2-1) absorption line with the IRAM millimeter
wave interferometer did not directly resolve the NE and SW components
\cite{wiklind98}. This is due to the low declination of the source
relative to the latitude of the interferometer, creating a
synthesized beam elongated in approximately the same direction
as the image separation. The continuum, however, is strong enough
to allow self-calibration, making it possible to accurately track
the phase center. The best angular resolution is achieved in right
ascension ($\sim 0\ffas1$) with a factor $\sim$2 worse resolution
in declination due to an elongated synthesized beam.
At frequencies outside the absorption line, the phase center should
fall on a line in between the NE and SW components. Assuming that the
flux ratio NE/SW is similar to that derived for longer wavelengths
($\sim 1.3-1.4$), the phase center should move towards positive RA
at frequencies where the absorption occurs. If the covering factor
is unity, $\Delta \alpha$ should be $\sim +0\ffas25$. 
This is exactly the amount of shift observed for the saturated
HCO$^+$(2-1) line (Fig.~\ref{pks1830spec1}a). A similar shift in
the declination of the phase center can also be seen and concurs
with these results \cite{wiklind98}. This result has also been
confirmed through BIMA observations where the two continuum components
have been separated \cite{swift01}.

\begin{figure}
\psfig{figure=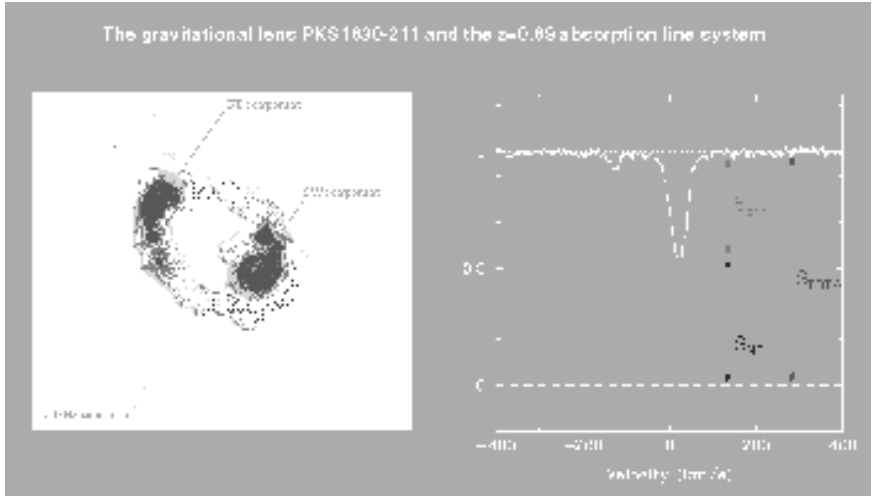,width=11.5cm,angle=0.0}
\caption[]{Illustration of how the individual light curves of the
NE and SW components can be derived from the single dish millimetric
observations. A 5 GHz radio image of PKS1830-211 \cite{subrahmanyan90}
shows the two cores and the extended jet emission. At millimeter
wavelengths only the cores contributes to the continuum emission. The
right hand panel shows the saturated HCO$^+$(2-1) spectrum. Since the
molecular gas only covers the SW component and the line opacity is
$\gg 1$, the depth of the absorption line corresponds to the flux from
the SW component only. The total continuum away from the absorption line
corresponds to the sum of the fluxes from the SW and NE components.}
\label{pks1830method}
\end{figure}

Due to this fortunate configuration of obscuring molecular gas in the
lensing galaxy, the flux contributions from the NE and SW cores
can easily be estimated using molecular absorption lines and a single
dish telescope with low angular resolution.
The 15m SEST telescope, which is used for the time delay monitoring
presented here, has a HPBW of $\sim$50$^{\prime \prime}$ at the observed
frequency of the HCO$^+$(2-1) transition, much larger than the
image separation of 0\ffas97.
Since molecular gas covers only the SW component, as shown by the
interferometric data, and the line opacity is $\gg 1$ as seen from the
rare isotopic lines, the depth of the absorption line corresponds to
the flux from the SW component only. The total continuum away from
the absorption line corresponds to the sum of fluxes from the SW and
NE components (Fig.~\ref{pks1830method}).

\subsubsection{Monitoring of HCO$^+$(2-1).}
\begin{figure}
\psfig{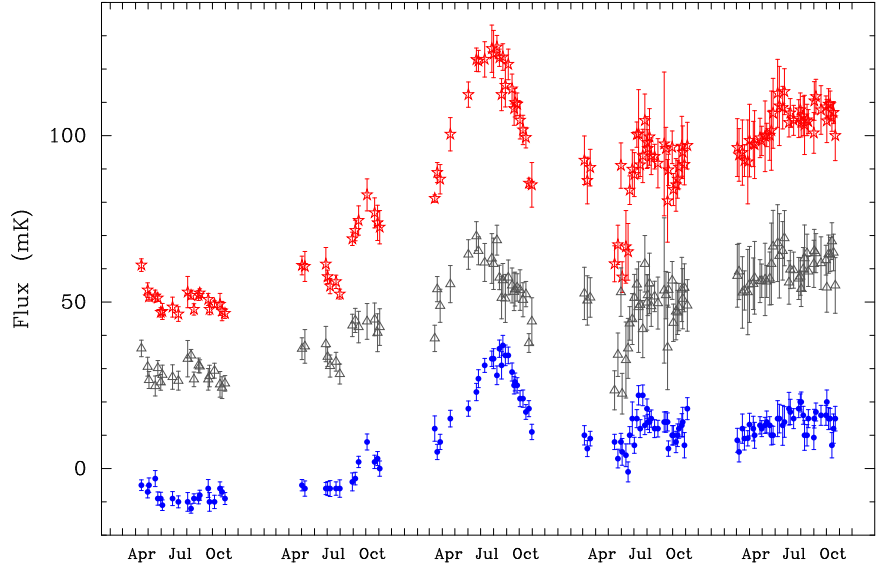}
\caption[]{Results from 5 years of monitoring of the HCO$^+$(2-1)
absorption at $z_{\rm d}=0.886$ towards PKS1830-211.
The top curve shows the measured total continuum flux away from
the absorption line. Notice the large outburst during 1998. The bottom
curve shows the depth of the HCO$^+$(2-1) line (shifted by
$-30$ mK). This corresponds to the flux from the SW component. The middle
curve shows the flux derived for the NE component.}
\label{pks1830monit1}
\end{figure}

\begin{figure}
\psfig{figure=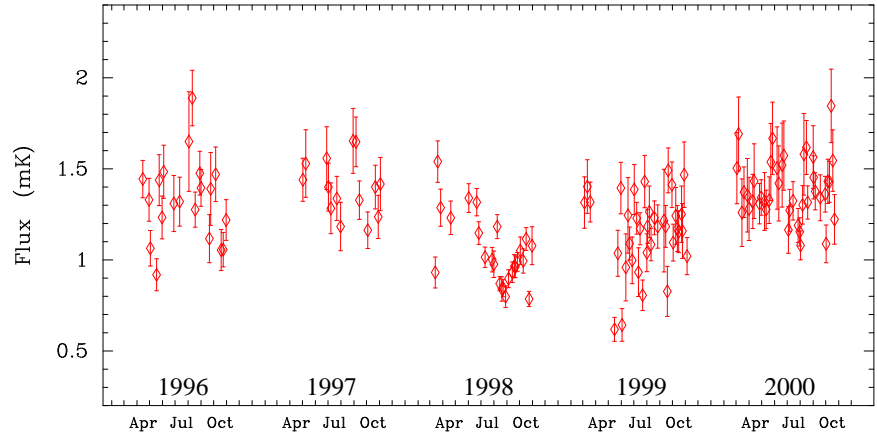,width=11.5cm,angle=-90.0}
\caption[]{Same as Fig.~\ref{pks1830monit1} but showing the flux
ratio of the NE and SW components.}
\label{pks1830monit2}
\end{figure}

\ 
Monitoring of the HCO$^+$(2-1) absorption and the total continuum
flux has been going on at the 15m SEST telescope since April 1996. Data
can only be obtained between February and November due to Sun constraints
(PKS1830-211 comes within 3$^{\circ}$ from the Sun). The light curves
are shown in Fig.~\ref{pks1830monit1}. Since only the total continuum
(at the top in the figure) and the depth of the absorption line (at the
bottom in the figure) are measured, the flux from the NE component (middle)
is derived as the difference $T_{\rm NE} = T_{\rm tot} - T_{\rm abs}$,
and therefore has a somewhat higher uncertainty.
During the 1996-2000 campaigns a total of 144 usable observations have
been obtained. The background quasar had a large outburst during 1998.
The outburst shows a single peak in the total continuum, putting an
upper limit to the differential time delay between the two cores (no
double peak structure). Separating the light curves for the two cores,
however, one can clearly see a delayed response of the SW image relative
to the NE (Fig.~\ref{pks1830monit1}). This is seen even more clearly when
plotting the ratios of the NE and SW fluxes (Fig.~\ref{pks1830monit2}).
This ratio shows the relative magnification of the two cores and should
be constant in the absence of a time delay. The decrease in the flux
ratio during the 1998 outburst is a clear indication of the time delay
between the NE and SW cores.

\subsubsection{Monitoring results.}\ 
The light curves shown in Figs.~\ref{pks1830monit1} and \ref{pks1830monit2}
have been analyzed using several different techniques.
The problem is straightforward: correlate two unevenly sampled time
series. However, the analysis is complicated by the fact that since
the two time series are really copies of each other but shifted in
time they are effectively sampled at different epochs.

There are two main types of methods used for obtaining the time delay.
One is to interpolate unobserved data points and then apply standard
techniques for cross correlation. The other is to use the unevenly
sampled time series and try to correlate neighboring points as good
as possible. The former method is superior if the sampling rate is
high, but this is usually not the case for astronomical data.
The latter method has less precision but is the least unbiased way
of obtaining the time delay. Both types of methods have been used
for the molecular absorption line data on PKS1830-211.

\subsubsection{Data analysis.}\ 
The observables at each epoch $t_i$ are the total continuum flux
$S_0(t_i)$ and the depth of the absorption line, corresponding to
the flux from the SW component $S_{SW}(t_i)$. The total continuum
is the sum of the two image components,
$$
S_0(t_i) = S_{1}(t_i) + S_{2}(t_i) \ \ ,
$$
where subscript 1 refers to the NE component and 2 to the SW component.
We know that the $S_1$ and $S_2$ fluxes are related as
$$
S_1(t_i) = \mu\,S_2(t_i + \Delta t) \ \ ,
$$
where $\mu$ is the magnification ratio and $\Delta t$ the differential
time delay between the two cores. Hence,
$$
S_0(t_i) = \mu\,S_2(t_i + \Delta t) + S_2(t_i)\ \ .
$$
By shifting the observed $S_2$ values, multiplying it with a magnification
ratio and adding the observed unshifted and unmagnified values we should
recover the observed total flux at time $t_i$. Since the observations 
consist of an unevenly sampled time series, with significant amount
of noise, finding the true $\Delta t$ and $\mu$ is a non-trivial exercise.

\medskip

The analysis of the light curves has been done using three methods.
Two of them involves interpolating between the observed data points
and construction of an evenly sampled time series. Due to the rather
long interruptions due to the Sun avoidance, interpolation only extends
over periods between February and November. An elaborate interpolation scheme
of unevenly sampled data has been developed with the specific goal of
resolving the time delay controversy of 0957+561 (Press et al.~\cite{press92}).
In the case of the molecular absorption data, however, a smoothing
function with an effective resolution similar to the average data point
separation at each epoch was applied. Using the smoothed time series,
data points in between observed epochs were linearly interpolated.
The smoothing dampens the worst fluctuations while retaining the small
scale structure in the time series. It also allows an easy assessment of
the relative weights of observed and interpolated data. A complication,
however, is that the data points are no longer completely independent.
This is of some concern when deriving the reduced $\chi^{2}$ values.

\subsubsection{$\chi^2$ minimization :} 
Minimization was done using both the time delay $\Delta t$ and the
magnification ratio $\mu$ as well as keeping the magnification ratio
fixed or time dependent. The latter is due to a surprising realization
that the magnification ratio might be variable, albeit on a
much longer time scale than the time delay (cf. Fig.~\ref{pks1830monit2}).
Using a parameterized $\mu(t_i)$ means that the solution becomes
cumbersome and slow. Instead we smoothed the flux ratio and fitted
a third order polynomial. This parameterization of $\mu(t_i)$ was
used when solving for $\Delta t$. The result, together with
results from the other analysis methods, is shown in
Fig.~\ref{pks1830monit3} and gives a $\Delta t = 27$ days.

\subsubsection{Cross correlation :} 
\begin{figure}
\psfig{figure=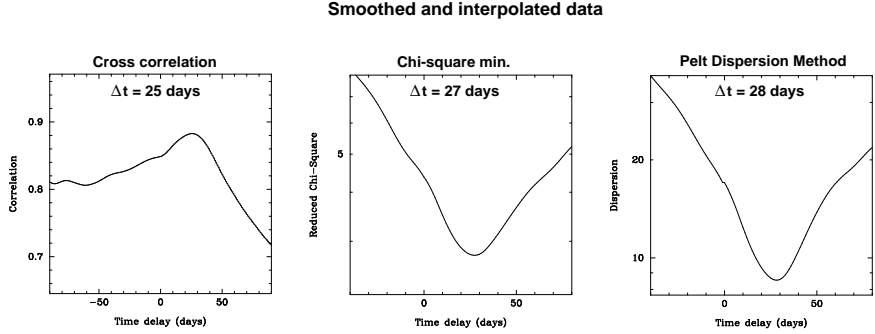,width=11.5cm,angle=-90.0}
\caption[]{The time delay for PKS1830-211 derived from molecular
absorption lines and using cross-correlation, $\chi^2$ minimization
and the Pelt Minimum Dispersion method. All three methods in this
example use the interpolated data set.}
\label{pks1830monit3}
\end{figure}

Edelson \& Krolik \cite{edelson88} developed a discrete cross correlation
method specifically aimed for reverberation mapping of AGNs that can be used
for time delays in gravitational lensing. The method optimizes the
binning of data points rather than the interpolation, as in the method
of Press et al.~\cite{press92}. The method requires a fairly well sampled 
data set to start with in order to retain a sufficiently good temporal
resolution. The sampling rate for molecular absorption line data in
PKS1830-211 is not dense enough to use the Edelson \& Krolik method.
Instead cross correlation was done on the same smoothed and interpolated
data set as the $\chi^2$ minimization. The cross correlation coefficient
is defined as $r_{\rm ab} = s_{\rm ab}^2/(s_{\rm a}\,s_{\rm b})$, where
the covariances $s_{\rm a}$ and $s_{\rm b}$ are defined in the usual manner
(cf.~\cite{bevington92}).
As with the $\chi^2$ minimization, the data points are not entirely
independent due to the smoothing and interpolation and the variances are
only approximately true. The result gives $\Delta t = 25$ days, with a
rather broad maximum for the cross correlation coefficient
(Fig.~\ref{pks1830monit3}).

\subsubsection{Minimum dispersion (the Pelt method) :}
\begin{figure}
\psfig{figure=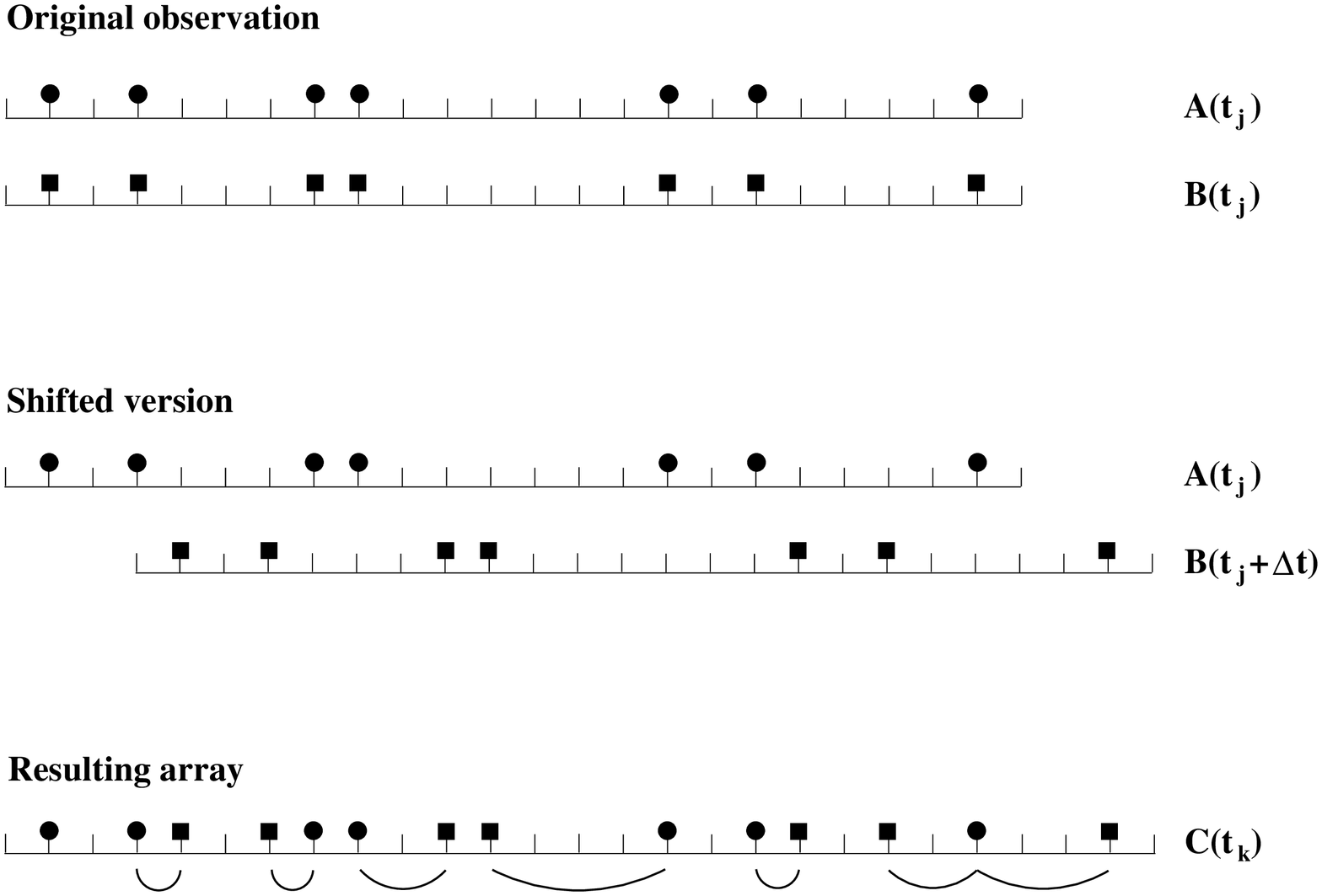,width=11.5cm,angle=-90.0}
\caption[]{Illustration of the Pelt Minimum Dispersion method used
to determine time delays in gravitational lenses. This method was
one of the methods used for deriving the time delay in PKS1830-211
from molecular absorption lines. This particular
case shows a two-image lens (A and B), where the respective
light curves are sampled at irregular intervals. In the middle
section the B light curve is shifted by $\Delta t$. By projecting
the resulting data points to a common array (C), nearest neighbors
of different light curves are correlated (arcs). A weight,
depending on the time difference between the points used in the
correlation, can be applied.}
\label{peltmethod}
\end{figure}

A simple and robust technique for analyzing unevenly sampled
time series was presented by Pelt et al.~\cite{pelt94} \cite{pelt96}.
They successfully applied it to the lens system 0957+561.
The strength of the method is that interpolation or smoothing
are not needed, leaving the errors for each data point independent.
The method is a form of cross correlation where a given data point
is correlated with a data point which is temporally its closest
neighbor. The method is illustrated in Fig.~\ref{peltmethod},
where the round and square markers in the two top rows represent
the two photometric data sets obtained from a two-component
gravitational lens. When correlating the time series, one of
them is shifted in time, as the square markers in the middle rows.
Projecting both the unshifted (round) and the shifted (square)
time series to a common array (bottom row), correlation is done
between those data points which are from different time series
and closest to each other. These points are connected by arcs
in the figure. It is easy to include the effects of different
magnifications for the lensed components as well as time delays
in systems with two or more lenses \cite{pelt98}.

\begin{figure}
\psfig{figure=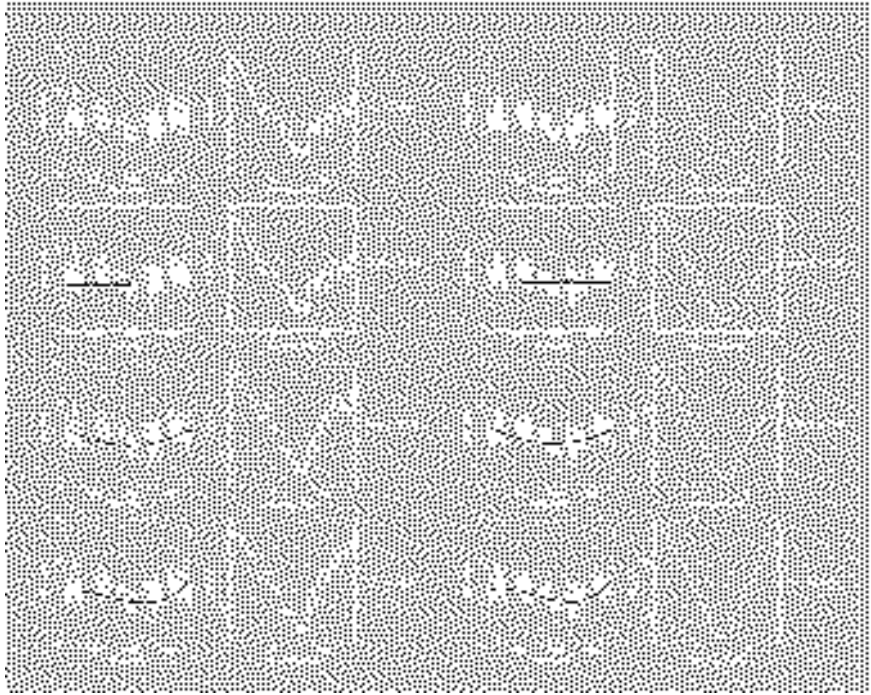,width=11.5cm,angle=0.0}
\caption[]{Results for the time delay of molecular absorption lines
in the PKS1830-211 gravitational lens system using the Pelt Minimum
Dispersion method. Both raw and interpolated data sets are shown
(left and right, respectively). Also shown are the results for
different treatments of the magnification ratio. In the top two
rows, the magnification ratio is included in the fit, while
in the two bottom rows the magnification ratio is predetermined by
fitting either a second or third degree polynomial to the observed
magnification ratio.}
\label{pks1830monit4}
\end{figure}

The results are undeniably noisier than for the interpolated data sets.
This can be seen in Fig.~\ref{pks1830monit4}, where the Pelt dispersion
method has been applied to both raw and interpolated data. The best
fit is for a time delay $\Delta t = 28$ days, with the NE component
leading.

\subsubsection{Error analysis :}
The errors associated with the light curves are a combination of noise
in the data points and systematic errors. The latter can originate in
the instrument, in the modeling necessary for separation of the lensed
components (as in the case of long wavelength radio observations),
assumptions made about the lensing system, etc. When interpreting
the time delay in terms of a Hubble constant, the largest systematic
error comes from modeling of the gravitational potential (see Chapter
X). Noise in the data comes from imprecise measurements but can also
originate in secondary variability such as microlensing and interstellar
scintillation. The latter is applicable at long radio wavelengths. 
Microlensing may be of importance even for gravitational lenses observed
at radio wavelengths (cf.~\cite{koopmans00}).

In order to assess the significance of correlations found in the light
curve of gravitationally lensed images it is customary to derive the
confidence limits through Monte Carlo simulations and bootstrap
techniques (cf.~\cite{fassnacht99}). The results often shows
non-Gaussian distributions and confidence levels are set by finding
the range of delays and magnification ratios inside which a given
amount (say 95\%) of the simulations lie. This gives a better
estimate of the true confidence level than simply fitting a Gaussian
to the distribution.
Doing this for the molecular absorption line data in PKS1830-211 gives
a time delay of $\Delta t = 28^{+4}_{-5}$ days, with the NE component
leading.

\begin{figure}
\psfig{figure=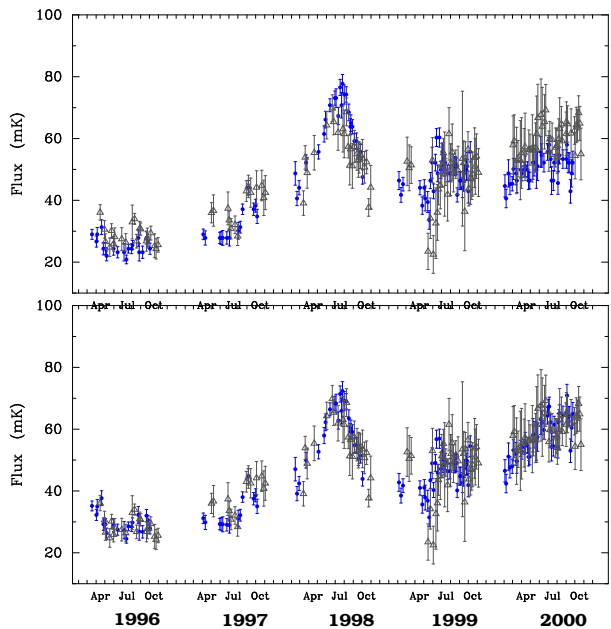,width=8cm,angle=0.0}
\caption[]{The light curve of the SW component in
PKS1830-211 (black) shifted by $\Delta t = -28$ days and
multiplied by a magnification ratio $\mu$. The light curve
for the NE component is shown in grey. In the upper panel
a constant magnification ratio of $\mu = 1.3$ was used,
while in the lower panel a time dependent magnification
ratio was used (cf. Fig.~\ref{pks1830monit4}).}
\label{pks1830monit5}
\end{figure}

In Fig.~\ref{pks1830monit5} the light curve of the SW component in
PKS1830-211 has been shifted by $-28$ days and multiplied by a
magnification ratio $\mu$. In the upper panel a constant ratio
of $\mu = 1.3$ was used, while in the lower panel a time dependent
magnification ratio was used. The use of different parameterizations
of the magnification ratios do not change the derived time delay,
but the time dependent form provides a better fit of the two light
curves. The reason for the slow change in magnification ratio is
presently unclear. It may have implications for the use of molecular
absorption lines as a probe of the time delay, but since the
time scale for the change of the magnification appears to be much
longer than the time delay, it is likely to be of small importance
when correlating light curves for each period (i.e. 9 months).

\section{LENS MODELS FOR PKS1830-211} \label{pks1830model}

\begin{figure}
\psfig{figure=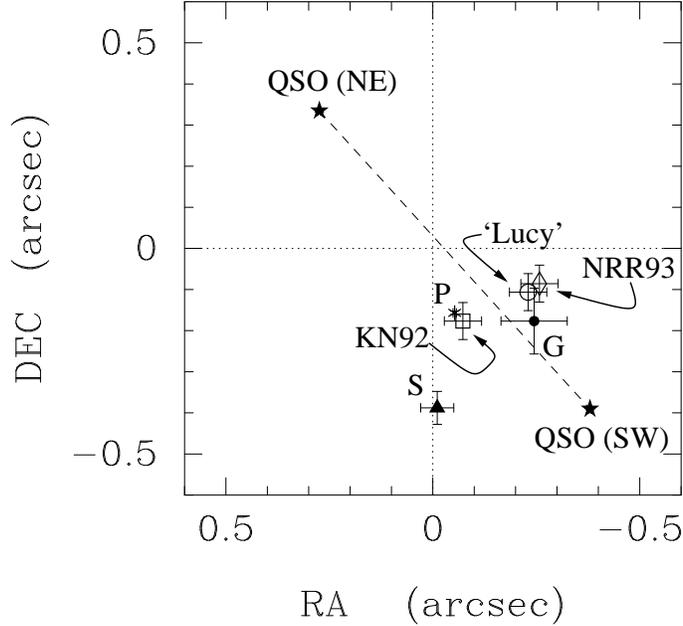,width=9.0cm,angle=0.0}
\caption[]{Illustration of the positions of various components in
the PKS1830-211 gravitational lens system. The two cores are marked
by QSO (NE) and QSO (SW). The putative location of the lens galaxy,
derived by Leh\'{a}r et al.~\cite{lehar00}, is marked by `G' (filled circle)
and a point source of unknown origin is marked by `P' (star).
A new possible lens location (from \cite{courbin02}) is marked by `S'
(filled triangle). The location of the center of the lens derived from
the lens model here, using the Lucy rectification scheme,
is marked by `Lucy' (open circle), the lens center derived by Nair et
al.~\cite{nair93} by `NRR93' (open diamond) and the lens center derived by
Kochanek \& Narayan \cite{kochanek92} by `KN92' (open rectangle).
The dashed line joining the two QSO images is only to guide the eye.}
\label{pks1830model_1}
\end{figure}

In order to use the differential time delay to derive a value for
the Hubble constant, a lens model has to be fitted to the observed
data. This is not a trivial exercise in most cases and this is
particularly true for PKS1830-211. Due to its location at Galactic
longitude $l=12.2^{\circ}$ and latitude $b=-5.7^{\circ}$,
PKS1830-211 suffers considerable Galactic extinction. In addition,
the molecular gas seen in absorption towards the SW component
contributes significant obscuration for at least this image.
Early attempts to identify the radio source PKS1830-211 with an optical
counterpart were all unsuccessful \cite{subrahmanyan90}
\cite{djorgovski92}. It was only with the advent of sensitive
infrared imaging and spectroscopical capabilities that progress could
be made. The NE image was positively identified using K-band imaging
at Keck and the ESO NTT \cite{courbin98}. While the redshift of the
lens, $z_{\rm d} = 0.886$, had been derived using molecular absorption
lines \cite{wiklind96a}, the redshift of the source was obtained
from near-infrared spectroscopy \cite{lidman98}. The redshift was
found to be $z_{\rm s} = 2.507$. Imaging with the HST WFPC2 and NICMOS
allowed identification of both the NE and SW image \cite{lehar00}.
In addition, an object which might be the lensing galaxy was detected
(designated as G). Its exact center position remains uncertain due to
the presence of a point source $\sim 190$\,mas away. The nature of the
point source remains unknown but could possibly be a Galactic star and
thus of no importance for the lens model.
In a recent paper by Courbin et al.~\cite{courbin02} combined images
from the HST and Gemini-North telescopes show what might actually be
the lensing galaxy at $z_{\rm d} = 0.889$. The lens has two spiral
arms, as expected from the molecular absorption data. One spiral arm
crosses the SW image of the QSO. The center of the spiral is, however,
significantly offset from the line joining the NE and SW images. Based
on symmetry arguments, the center of the lensing galaxy is believed to
be in the proximity of this line joining the images (Fig~\ref{pks1830model_1}).

All the necessary ingredients for a detailed lens model are thus in
place, except for two remaining uncertainties: the exact position of
the lensing galaxy and the possible double-lens nature of the system
(cf. Sect.~\ref{molabs}).

\subsection{Early models}

There have been several attempts to model the lens system PKS1830-211.
The system was first detected at radio wavelengths, where it is a
prominent southern radio source. The morphology of the system was found
to be that of a double, while the radio spectrum is typical for
a compact flat-spectrum source. This led Rao \& Subrahmanyan \cite{rao88}
to first suggest that PKS1830-211 is a gravitationally lensed system.
Based only on the radio images and their polarization properties, obtained
with the Very Large Array (VLA) at 5 and 15 GHz, Subrahmanyan et al.~\cite{subrahmanyan90}
constructed a lens model which is not much different from later ones based
on more detailed data. In order to reconstruct the extended radio structure,
Subrahmanyan et al. modeled the source as an one-sided core-jet structure.
To get lensed images with a morphology similar to the observed one, they
also had to include a `knot' in the jet. Based on their lens model
Subrahmanyan et al. predicted a time delay of $27\,h_{100}^{-1}$ days
(using the now known redshift of the source and the lens).

Nair et al.~\cite{nair93} modeled the PKS1830-211 system using improved radio
interferometry data (cf.~\cite{jauncey91}). The method was similar
to that of Subrahmanyan et al.~\cite{subrahmanyan90} in that the source structure was
built up in a piecemeal manner in order to fit various observed features.
With this type of method one can emphasize the influence of small and weak
features which may carry a small weight in an inversion scheme
based on $\chi^2$ minimization of model$-$observed results, but may
nevertheless carry important information on the lensing scenario.
In the PKS1830-211 system such a weak radio feature, labeled E, was used
by Nair et al. to constrain the lens model. This feature might be
a third demagnified image of the core. However, since the flux of the
E component is less than one percent of the peak value, its significance
in terms of flux is small unless the dynamic range of the interferometry
maps is very good. Nair et al. found that an elliptical potential with
the radio core located close to the inner edge of the radial caustic
gave a good fit to the observed morphology. As in the previous model
by Subrahmanyan et al.~\cite{subrahmanyan90}, it was necessary to include
a `knot' feature in the source distribution.
The jet needed to be bent and cross the tangential caustic. The model
gives a good fit to the observed system and places the lens galaxy close
to, but not coinciding with, the possible lens position observed by
Leh\'{a}r et al.~\cite{lehar00}. The estimated time delay, using the known source
and lens redshifts, is $17\,h_{100}^{-1}$ days.

\bigskip

A difficulty with extended lensed images is that any inversion must solve
simultaneously for both the lens configuration and the source structure. 
This type of inversion problem can be seen as
\begin{eqnarray} \label{eq1}
I_{obs}(\zeta) & = & \int{\psi(\zeta^{\prime})\,K(\zeta - \zeta^{\prime})}\,
d\zeta^{\prime}\ \ ,
\end{eqnarray}
where both the source distribution $\psi(\zeta)$ and the kernel $K$ (here
representing the lensing potential) are unknown.
This type of problem is generally unsolvable. In the case of lensing, however,
one can use the knowledge that when the lensed image contains multiple
distorted components of the background object these must arise from a
common source. Furthermore, it is known that surface brightness is conserved.
These `priors' constrain the problem and permit the simultaneous solution
of both the structure of the source and the properties of the lensing potential.
This type of inversion problems for gravitational lenses has been developed
extensively by Kochanek, Wallington and collaborators in several papers
(cf.~\cite{kochanek89} \cite{kochanek92} \cite{wallington96}).
In particular, Kochanek \& Narayan \cite{kochanek92} developed an inversion method
based on the CLEAN routine (cf.~\cite{hogbom74}) used in radio interferometry
data reduction and applied it to PKS1830-211. This method takes into
consideration the effects of the finite resolution when attempting to
invert the lens model and is thereby able to better distinguish the best
lens model. 
The LensClean method of Kochanek \& Narayan has produced the hitherto most
reliable model for the PKS1830-211 system, but due to the finite resolution,
the inversion was done on radio data with rather low angular resolution
but with good signal-to-noise, it is not likely to represent the final model.

Leh\'{a}r et al.~\cite{lehar00} modeled the PKS1830-211 system using a singular
isothermal elliptical mass distribution as well as with two singular isothermal
spheres representing the lens galaxy and the source G2 (cf. Sect.~\ref{molabs}).
In both these cases they fixed the lens at the position of G, with a positional
uncertainty of 80\,mas. The extended radio emission was not used to constrain
the lens model. Leh\'{a}r et al. noted the strong dependence of the location
of the lens galaxy and the Hubble constant derived from differential time delay
measurements.     

\begin{figure}
\psfig{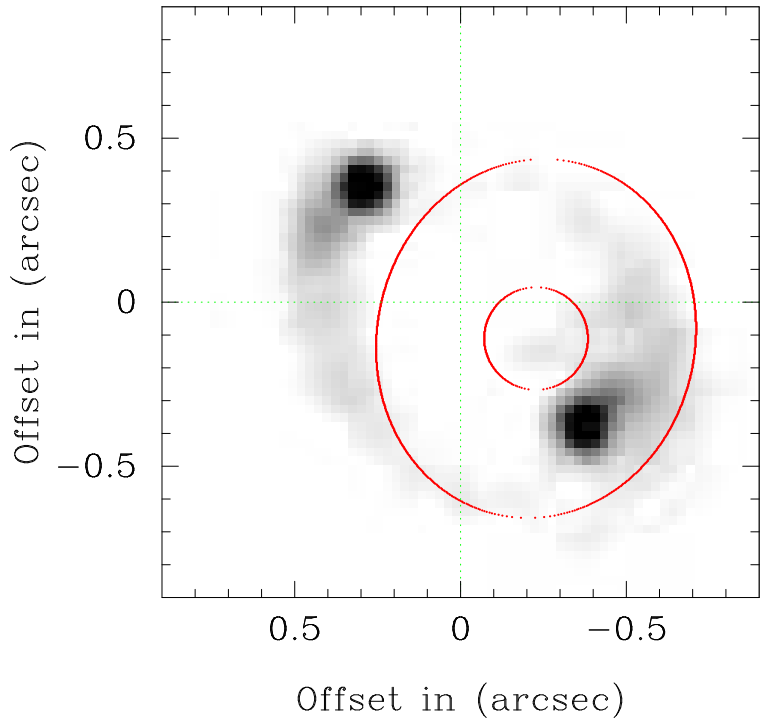}
\caption[]{A 15 GHz radio image of PKS1830-211 (from \cite{subrahmanyan90}).
The center of the coordinate system is arbitrary. All
positions in the text and in Table~\ref{lensmodel-table} are relative
the NE image. The critical lines of the best fit lens model, shown
in Fig.~\ref{lensmodel-fig2} are shown for comparison.}
\label{lensmodel-fig1}
\end{figure}    

\begin{figure}
\psfig{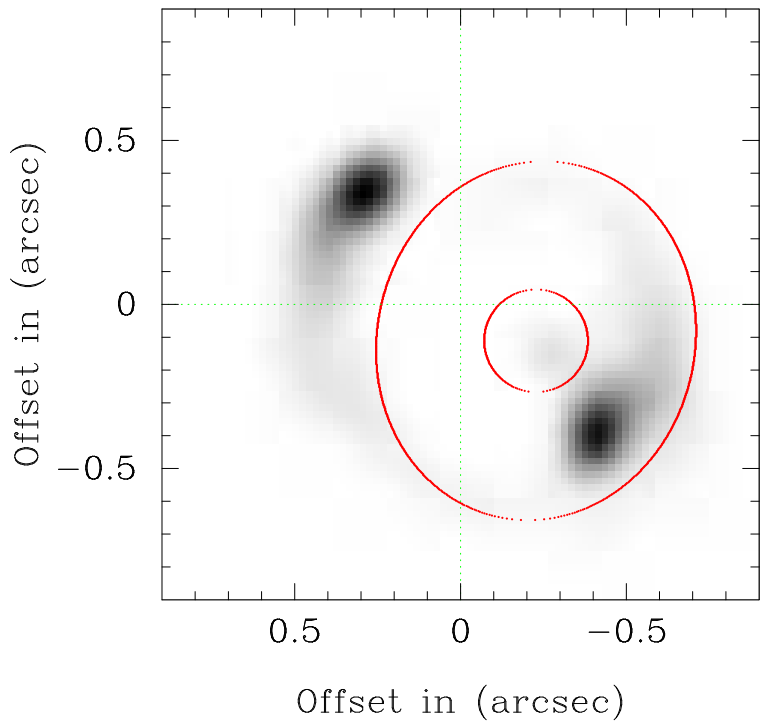}
\caption[]{The lensed images obtained from inversion of the lens
equation  using Lucy rectification and Simulated Annealing (as
described in the text). In this particular solution, the lens
is fixed at the position of the observed (putative) lens center
$G$ (see Fig.~\ref{pks1830model_1}). The critical lines from the
lens model are marked.}
\label{lensmodel-fig2}
\end{figure}

\begin{figure}
\psfig{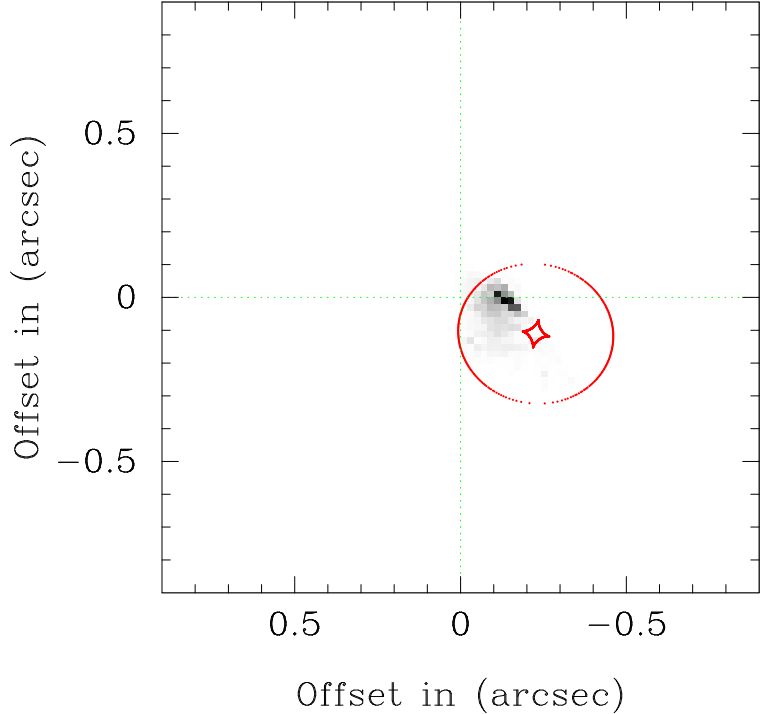}
\caption[]{The source distribution derived from the best fit lens
model, which gives the image seen in Fig.~\ref{lensmodel-fig2}. The
caustic structure is shown.}
\label{lensmodel-fig3}
\end{figure}

\subsection{A new lens model of PKS1830-211}

The situation is rather unsatisfactory concerning the various solutions
to the lensing configuration characterizing PKS1830-211. Three different
models (\cite{kochanek92} \cite{nair93} \cite{lehar00})
give three different positions for the lens, with corresponding differences
in the value of the Hubble constant for a given time delay. The situation
is summarized in Fig.~\ref{pks1830model_1}.
The location of the cores in the long wavelength radio data used by Kochanek \&
Narayan \cite{kochanek92} differ from that used by others (using data from shorter
wavelengths). Their lens center relative to the other features shown in
Fig.~\ref{pks1830model_1} is therefore somewhat uncertain.

\medskip

The fact that surface brightness is conserved can be used in modeling a
lensing configuration if both the observations and the code have infinitely
good resolution. In the opposite extreme, if the source remains unresolved,
one can calculate the magnification for a point source and apply a
smoothing function (convolution) representing the observational transfer
function (e.g. atmosphere, telescope, imaging array). A more difficult
situation arises when an extended source distribution is partially resolved
by the observer. 
A given resolution element will represent different areas of the source in
a rather complicated manner and the conservation of surface brightness
ceases to be a good prior for an observed lens system. In their LensClean
method Kochanek \& Narayan \cite{kochanek92} solves this situation by representing
the source emissivity distribution by $\delta$-functions, mapped through
the lens configuration with the corresponding magnification `turned on'
and then smoothed by a restoring beam.

An alternative way to solve the inversion problem as stated above is
presented here. A more thorough description of the method will be published
in Wiklind (2002). The concept is similar to the LensClean method of Kochanek
\& Narayan \cite{kochanek92}. However, instead of introducing CLEAN components,
as $\delta$-functions to represent the source distribution, this method starts
with a none-zero smooth source distribution and applies the Lucy rectification
method \cite{lucy74} to constrain the source emissivity, given the observed
lensed images and for a given lens model.
The Lucy rectification scheme has been used extensively to
deconvolve images obtained with the Hubble Space Telescope before the
corrective optics was installed. It has also been used to deconvolve
the molecular gas distribution in galaxies observed with single dish
telescopes of rather poor angular resolution \cite{wiklind92}
\cite{wiklind93}.

\medskip

The method has an outer loop, which controls the lensing parameters,
and an inner loop which solves for the best source distribution
given the lens configuration. In the inner loop the source distribution
is mapped through the lens and the resulting image is compared with the
observed one. The source distribution is adjusted according to the Lucy
method (see below) and the process is repeated. When the inner loop has
converged, the source distribution is mapped through the lens a final
time and the resulting image gives a goodness-of-fit. The lensing
configuration is then modified in the outer loop and the process is repeated.
The Lucy method is used only in the inner loop, each time starting with a
perfectly smooth source distribution.
The observed images are represented by
\begin{eqnarray} \label{eq_lucy1}
I_{obs}(x_1,x_2) & = & \int{\int{\psi(y_1,y_2)\,
K(x_1,x_2|y_1,y_2)\,dy_1dy_2}}\ \ ,
\end{eqnarray}
where $\psi(y_1,y_2)$ is the true source distribution and $K(x_1,x_2|y_1,y_2)$
is the kernel representing both the gravitational lens and the finite angular
resolution of the observation. The kernel is here written as a conditional
probability function: the likelihood of $(x_1,x_2)$ given $(y_1,y_2)$.
With this formulation we can use the Lucy method in a straightforward manner.
The idea being that an approximation to the true source distribution is
\begin{eqnarray} \label{eq_lucy2}
\psi^{n+1}(y_1,y_2) & = & \psi^{n}(y_1,y_2)\,
\int{\int{\frac{I_{obs}(x_1,x_2)}{I^{n}(x_1,x_2)}\,K(x_1,x_2|y_1,y_2)\,
dx_1dx_2}}.
\end{eqnarray}
The conditional probability function $K$ contains the likelihood of an image
emissivity at position $(x_1,x_2)$ given a source emissivity at $(y_1,y_2)$
and the action of a restoring beam (i.e. a finite angular resolution). The
simplest (although not entirely correct) restoring beam is a Gaussian with
a HPBW similar to the angular resolution of the observations.

Even starting with a constant source emissivity distribution $\psi^{0}$,
the Lucy rectification converges very rapidly to a specific source distribution.
Unfortunately, there is no good criteria for determining when to stop the
rectification (cf.~\cite{lucy74}). In this particular application the inner
loop was stopped after 5 iterations of the Lucy rectification. The outer
loop consists only of changing the lensing parameters. Several different
methods can be employed for this, the most efficient for this application
being the Simplex method. However, there is a risk that this method gets
stuck in a local minimum and great care has to be taken to ensure that
a global minimum has really been reached. This involves repeatedly
restarting the Simplex method with parameters offset from the ones
giving a (local?) minimum in the $\chi^{2}_{\nu}$. An alternative method that
circumvents this, but that is computationally more expensive, is Simulated
Annealing (cf.~\cite{metropolis53} \cite{wiklind92}).
This latter method was used in this application.

\medskip

The lens was modeled as a non-singular elliptical mass
distribution
\begin{eqnarray} \label{massmodel}
\kappa(x_1,x_2) & = & \kappa_{0}\,\left(x_{1}^{2} + \frac{x_{2}^{2}}{q^2}
+ s^2\right)^{-\gamma}\ \ ,
\end{eqnarray}
where $q$ is the projected axis ratio, $s$ is the core radius and the
surface density profile is set to $\gamma = 1/2$, representing an
isothermal mass distribution. The deflection angle and magnification
was calculated using the code developed by Barkana~\cite{barkana98}.
This code can handle surface density profiles with $\gamma \neq 1/2$,
but here the modeling is restricted to the isothermal case.
The density profile is, however, a very important parameter when
deriving the Hubble constant using differential time delays
(cf.~\cite{williams00} \cite{koopmans99}).
In all there are six lens parameters that were
fitted: the center position of the lens, the position angle, the
velocity dispersion, the ellipticity, and the core radius. 

Applying this method to the 15 GHz radio image of PKS1830-211 shown
in Fig.~\ref{lensmodel-fig1} \cite{subrahmanyan90}, a best
fit lens model is achieved with the parameters as tabulated in
Table~\ref{lensmodel-table}. Also listed in Table~\ref{lensmodel-table}
are the results when keeping the lens position fixed at the coordinates
of the putative infrared lens center G.
The resulting image distribution for case (b) is seen in
Fig.~\ref{lensmodel-fig2} together with the critical lines of the
lens. The corresponding source distribution is seen in
Fig.~\ref{lensmodel-fig3} together with the caustic structure.

The two solutions presented in Table~\ref{lensmodel-table} are very
similar to each other, yet they give quite different values to the
Hubble constant. Using the lens model of Nair et al.~\cite{nair93}, the
Hubble constant becomes H$_{0} = 59^{+11}_{-8}$ km/s/Mpc.
These differences are mainly due to the different locations of
the lens center and introduces a large uncertainty in the correct
value of H$_0$.

\medskip

The lens model presented here will be further refined and the results should
be regarded as tentative. However, unless the position of the lens can
be derived more accurately, the value of the Hubble constant will remain
uncertain. As mentioned above, the shape of the density profile is also
a source of uncertainty for a more exact derivation of H$_0$.
This  uncertainty is largest for exponents $\gamma < 1/2$ \cite{koopmans99}.

\begin{table}
\caption[]{Lens parameters}
\begin{center}
\renewcommand{\arraystretch}{1.4}
\setlength\tabcolsep{5pt}
\begin{tabular}{c|rr|rr}
\noalign{\smallskip}
\hline
\noalign{\smallskip}
\multicolumn{1}{c|}{}                                &
\multicolumn{2}{c|}{6 free parameters}               &
\multicolumn{2}{c}{4 free parameters}                \\
\multicolumn{1}{c|}{}                                &
\multicolumn{2}{c|}{}                                &
\multicolumn{2}{c}{(lens fixed)}                     \\
\noalign{\smallskip}
\hline
\noalign{\smallskip}
lens center$^{a}$  & -0.5008 & -0.5205 & -0.5010 & -0.4450$^{b}$ \\
$\kappa_{0}$ & 0.3196  &         & 0.3106  &         \\
$q$          & 0.8576  &         & 0.8991  &         \\
$s$          & 0.0929  &         & 0.0745  &         \\
PA           & 135.553 &         & 101.174 &         \\
\noalign{\smallskip}
\hline
\noalign{\smallskip}
$\left(\frac{\Delta t}{28^{\rm days}}\right)^{-1}$H$_{0}^{c}$ &
$63^{+14}_{-6}$ &         & $83^{+18}_{-9}$ &        \\
\noalign{\smallskip}
\hline
\noalign{\smallskip}
\end{tabular}
\end{center}
\ \\
$(a)$\ Relative to the NE component. \\
$(b)$\ Fixed at possible lens center G \cite{lehar00}. \\
$(c)$\ With $\Delta t = 28^{+3}_{-5}$ days.
\label{lensmodel-table}
\end{table}

\section{FUTURE PROSPECTS} \label{futureprosp}

Existing millimeter and submillimeter telescopes use gravitational
lenses more as an aid to the study of distant objects, rather than 
being an aid to the study of gravitational lensing as such.
Nevertheless, some information about the content of the interstellar
medium in both lenses and sources have been obtained, and the
molecular absorption lines seen towards PKS1830-211 have been used
to measure the differential time delay in this particular system.

This situation will change dramatically when planned telescopes at
millimeter and submillimeter wavelengths become available. Increased
sensitivity and angular resolution will make this wavelength regime
very important for studies of gravitational lenses as a phenomenon
of their own. The most obvious advantage is that obscuration effects
will be completely absent. The effects of microlensing will also be
absent or at least minimal. The use of flux ratios of lensed components
for constraining parameters when modeling lensed have fallen out of
favor due to differential extinction and microlensing effects, but
will be usable when the new submm/mm instruments are available.

Existing submillimeter and millimeter facilities include single dish
telescopes, such as the IRAM 30m telescope on Pico Veleta in Spain, the
JCMT 15m telescope on Mauna Kea, and the SEST 15m telescope on La Silla in
Chile. Two 10m size dishes, aimed primarily for submillimeter wavelengths,
include the CSO on Mauna Kea and the HHT on Mount Graham in Arizona.
These telescopes use both heterodyne receivers for spectral line observations
and bolometer type array cameras for continuum observations. The sensitivity
depends largely on the quality of the site and the instrumentation. The
angular resolution, however, is determined by the diffraction limit of the
telescopes. At $\lambda=1$mm, the angular resolution is limited to
$10^{\prime\prime}-25^{\prime\prime}$. This constitutes the largest limitation
to the study of gravitational lenses.
For number counts (see Sect.~\ref{numbercounts}) the lack of angular resolution
means that with only slightly more sensitive receivers, confusion will become
a major limitation (cf.~\cite{hogg01})

A few interferometers operating at millimeter wavelengths exist. The IRAM
Plateau de Bure interferometer in France consists of five (soon to be six)
15m telescopes, and represents the largest collecting area today. The OVRO
interferometer consists of six 10m telescopes, while BIMA consists of eight
6m telescopes.
In Japan the Nobeyama interferometer consists of six 10m telescopes. All of
these facilities operate at $\lambda=3-1$mm. The angular resolution reached
is typically around 1$^{\prime\prime}$ or slightly better. However, sensitivity
becomes a serious limitation at the longest baselines and highest angular
resolutions. Also, the Australian Telescope Compact Array (ATCA) has recently
been upgraded to work at 3mm with five of its 22m elements.

\begin{figure}
\psfig{figure=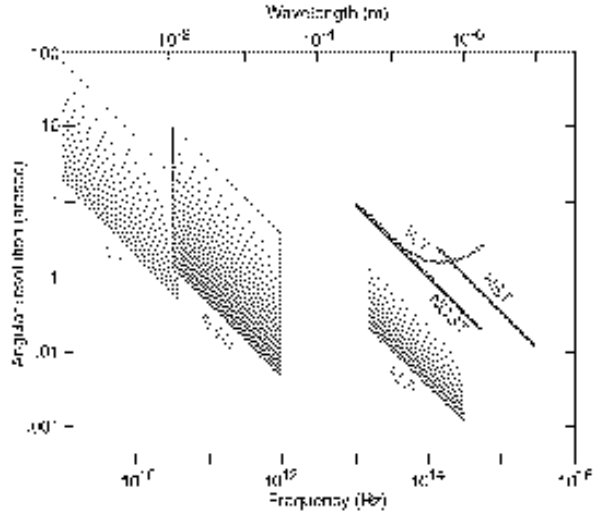,width=11.5cm,angle=-90.0}
\caption[]{A comparison of the wavelength coverage and the angular resolution
of existing and planned instruments. The VLA refers to the Extended VLA. With
ALMA, the same angular resolution will be reached from cm wavelengths into the
optical.}
\label{almafig1}
\end{figure}

\subsection{Future instruments} \label{future}

\subsubsection{Single dish telescopes.}\ 
A few single dish submm/mm telescopes are under construction, or in
advanced planning. These include the recently commissioned Green Bank
Telescope (GBT), which has a 90m unobstructed dish and will reach
$\lambda$3mm when fully operational. The Large Millimeter Telescope
is a 50m telescope built as a collaboration between INAOE and the University
of Massachusetts in Amherst. This telescope will be placed on Sierra Negra
in Mexico and operate at a wavelength of 1-4 millimeter.
Two single dish telescopes to be situated close to the ALMA site (see below)
are under construction:
APEX is a 12m, single dish telescope which will be placed on Chajnantor
at an altitude of 5000m. This telescope
will operate into the THz regime, i.e. $\lambda \sim 300\mu$m. ASTE is
a Japanese 10m dish, to be placed at Pampa La Bola, a few kilometers away
from the ALMA site. The ASTE will also operate at submm to THz frequencies.
These new single dish telescopes will explore distant objects, including
gravitational lenses, with somewhat better angular resolution than existing
telescopes. Nevertheless, even at the highest frequencies it will only reach
an angular resolution of $\sim 6^{\prime\prime}$. This is insufficient for
detailed studies of gravitationally lensed systems.

\subsubsection{The Atacama Large Millimeter Array.}\ 
A major step in submm/mm wave instruments will be the joint
European-US project, with Japanese involvement as well, of building
a large millimeter and submillimeter interferometer at an altitude of
5000m on Chajnantor in Chile.
This instrument, with the acronym ALMA, will consist of $64 \times 12$m
telescopes, each with a surface accuracy of at least 20$\mu$m. The total
collecting area will be 7238 m$^2$, which is an order of magnitude greater
than the largest existing instrument today. The longest baseline will be
$10-12$ km, leading to an angular resolution surpassing that of the
Hubble Space Telescope. A rough estimate of the angular resolution power is 
$0\ffas2\,\lambda_{\rm mm}/{\rm L}_{\rm km}$, where $\lambda$ is the wavelength
in millimeters and $L$ is the baseline length in kilometers. ALMA will easily
reach 0\ffas1 resolution and, at the highest frequencies and longest baselines,
0\ffas01. A comparison of the wavelength coverage and the projected angular
resolution for existing and planned telescopes is shown in Fig.~\ref{almafig1}.

At the same time ALMA will increase the sensitivity over existing
instruments by at least two orders of magnitude. The noise rms level
reached with an interferometer consisting of $N$ array elements, each
of effective area $A_{\rm eff}$ and with an integration time of $t_{\rm int}$
seconds over an effective bandwidth of $B$ Hz, is expressed as
\begin{eqnarray} \label{system1}
\Delta S & = & \frac{2k}{A_{\rm eff}}\,\frac{T^{\prime}_{\rm sys}\,e^{\tau_{\nu}}}
{\sqrt{N(N-1)\,B\,t_{\rm int}}}\ \ ,
\end{eqnarray}
where the term $e^{\tau}$ represents the damping by the atmosphere at frequency $\nu$.
The system temperature, $T^{\prime}_{\rm sys}$, is the total noise received by the
telescope (including ground pick-up).
From this expression it is clear that large individual antenna sizes,
a large number of array elements, a large bandwidth, a low system temperature
and as little damping from the atmosphere are essential for a sensitive interferometer.
These criteria can be fulfilled by using state-of-the-art receiver systems and putting
them at a high and dry site such as Chajnantor in the Andes.

The excellent atmospheric conditions at Chajnantor is estimated to give a total
system temperature (including the atmospheric damping) of $\sim$230 K at a wavelength
of 0.8mm. Using a 16 GHz wide backend and one hour of integration, the 5$\sigma$
detection limit will be 100\,$\mu$Jy. At a wavelength of 460\,$\mu$m the corresponding
5$\sigma$ detection limit after one hour is 1\,Jy. Although considerably worse, this
wavelength range is usually not reachable at all with ground-based facilities.
These values are representative for unresolved sources. If the source is resolved
at the longest baselines, the sensitivity decreases.

\begin{figure}
\psfig{figure=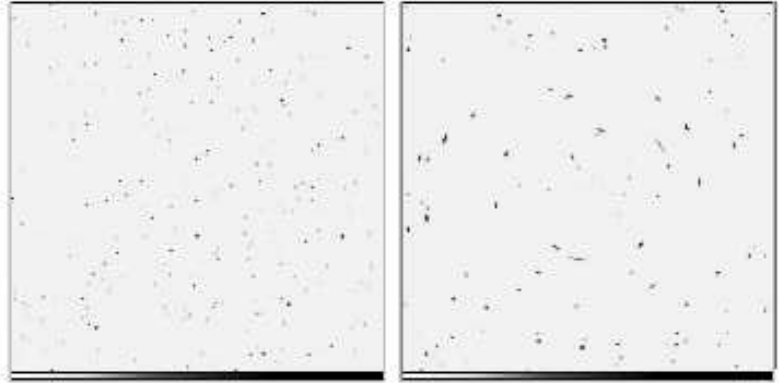,width=11.5cm,angle=-90.0}
\caption[]{A simulated 9 arcmin$^2$ deep field observed with ALMA at a wavelength
of 850$\mu$m. All sources above the 5$\sigma$ noise rms limit of 0.15 mJy are
shown. The source population was modeled to fit the observed cumulative source
counts, extrapolated to 0.1 mJy using a more shallow powerlaw than what can be
fitted in the range 1-10 mJy. {\bf Left:} Unlensed field. Approximately 210
sources are `detected'. {\bf Right:} The same field as in the left image but
with an intervening rich galaxy cluster at redshift $z=0.3$.}
\label{almafig2}
\end{figure}

\subsection{Weak lensing at submillimeter wavelengths}

With its superb sensitivity and angular resolution, ALMA can successfully
be used to study weak lensing by intermediate redshift galaxy clusters.
Assuming a constant co-moving volume density of dusty high redshift objects,
the constant sensitivity to dust continuum emission from $z\approx 1$ to
$z \approx 10$ (Sect.~\ref{dustdetect}) favors the detection of the highest
redshift objects. This selection bias is unique for the submm/mm wavelength
regime.
Furthermore, a reasonably well-sampled $uv$-plane, resulting from a large
number of interferometer elements, will have a point-spread function (PSF)
which is well behaved in comparison to existing wide-field CCDs. And, finally,
the dust distribution in galaxies tend to be symmetric and centrally
concentrated, leading to a simple geometry of the lensed sources.
The latter issue is, of course, not well constrained by existing data.
Although dust continuum emission will be easier to detect at large distances
than line emission (cf.~\cite{combes99b}), the high sensitivity and
broad instantaneous bandwidth of ALMA will allow CO lines to be observed
for many of the lensed sources. This enables a direct way of deriving the
redshift distribution of the source population. Furthermore, observations
of CO lines in the intervening galaxies will allow a determination of the
dynamical mass of the lenses.

ALMA will, however, have a relatively limited instantaneous field of view.
This is particularly aggravating at high frequencies. The field of view
is limited by the size of the individual telescopes and will be similar
in size to that of existing single dish telescopes. This means that the
time consuming process of mosaicing is necessary. The key parameters for
the effectiveness or speed of this new instrument are the sensitivity
$\Delta S$, the angular resolution and the primary beam area $A_{\rm fov}$.
The time required to survey a given area $A$ to a flux density limit $S$
is then: $t \approx (\Delta S/S)^{2}(A/A_{\rm fov})$ (cf.~\cite{blain99d}
\cite{blain96}).

An example of what can be achieved with ALMA is shown in Fig.~\ref{almafig2}.
The left panel shows a simulated 9 arcmin$^2$ field observed with ALMA to a
5$\sigma$ noise rms limit of 0.15 mJy. The wavelength is 0.8mm and was
chosen to maximize the number of detected dust continuum sources while
minimizing the observing time (by making the field-of-view as large as
possible). The field of view per pointing is $\sim$0.07 arcmin$^2$.
As shown in Sect.~\ref{future} it takes somewhat less than one hour
of telescope time to reach this 5$\sigma$ detection limit.
To cover 9 arcmin$^2$ thus requires a total of 130 hours. In this
example more than 200 sources will be detected (left panel) and their
distortion by a rich cluster at $z=0.3$ will be clearly detectable
(right panel).
In this example a population of dusty galaxies was assumed to have a
constant co-moving volume density between $z=1-7$ and undergo pure
luminosity evolution. The observed flux densities were fitted to the
number counts derived from SCUBA and MAMBO observations
(Sect.~\ref{numbercounts}). The median luminosity of the `detected'
galaxies is $3 \times 10^{11}$ L$_{\odot}$ and their redshift distribution
is more or less flat between $z=2$ to $z=7$. The cumulative source counts
was not, however, extrapolated to weaker flux densities using the same
powerlaw as applicable between 10-1 mJy (cf. Fig.~\ref{counts}, but with
a more modest slope. This ensures that the number of sources seen in the
right panel of Fig.~\ref{almafig2} is not an overestimate.

\medskip

In the same manner as lensing by intermediate redshift clusters is presently
being used to observe dust continuum sources at flux density levels otherwise
not reachable with existing instruments, ALMA will reach flux densities
approaching $\mu$Jy levels when observing lensed sources. For high redshift
galaxies, this flux density level corresponds to luminosities around $10^9$
L$_{\odot}$, i.e. dwarf galaxies.

\section{SUMMARY} \label{summary}

The study of gas and dust at high redshift is important for several reasons.
It gives us an unbiased view of star formation activity in obscured objects
and it tells the story of the chemical evolution and star formation history
in galaxies through the amount of processed gas (and dust) it contains.
With today's millimeter
and submillimeter facilities, this research area has used gravitational lensing
mostly as a tool to boost the sensitivity. This is evident through the preponderance
of gravitationally lensed objects among those which have been detected at $z > 2$
in the lines of the CO molecule. It is also evident in the use of lensing
magnification by galaxy clusters in order to reach faint submm/mm continuum
sources. There are, however, a few cases where millimeter lines have been
directly involved in understanding lensing configurations.
The best example of this is the highly obscured PKS1830-211, where the lens was
identified through molecular absorption lines and where these lines give a
velocity dispersion measure by originating in two different regions of the lens.
The molecular absorption lines in this system have also been used to derive the
differential time delay between the two main components, the main objective
being to determine the Hubble constant, but also adding to the constraints
in modeling this particular lens system.

With future millimeter and submillimeter instruments, such as ALMA, coming
on-line, the situation is likely to change drastically. The sensitivity of
ALMA will be such that it does not need the extra magnification from lensing
to observe very distant objects. Instead it will be used to study the lensing
itself. The more or less constant sensitivity to dust emission over a redshift
range stretching from $z \approx 1$ to $z \approx 10$ means that the likelihood
for strong lensing of dust continuum detected sources is much larger than
for optically selected sources. ALMA will therefore discover many more lenses
and allow a direct assessment of cosmological parameters through lens statistics.
Weak lensing will also be an area where ALMA can successfully contribute. Again,
the high sensitivity to dust emission out to very high redshifts, combined with
an angular resolution $< 0\ffas1$, and a more beneficial `PSF' will
make ALMA more efficient for probing the potential of galaxy clusters than
present day optical/IR telescopes. In addition we will be able to study both
the sources and the lenses themselves, free of obscuration and extinction
corrections, derive rotation curves for the lenses, their orientation and,
thus, greatly constrain lens models.

\bigskip

\noindent
{\bf Acknowledgments.}\ \ 
T.W. thanks Fran\c{c}oise Combes for allowing the use of unpublished material
on millimeter wave absorption line systems. Many thanks to F. Combes, D. de
Mello, P. Cox and F. Courbin for careful reading of the manuscript and for
valuable comments.

%

\end{document}